\documentclass[%
 aip,
 jmp,%
 amsmath,amssymb,
preprint,%
]{revtex4-1}

\usepackage{graphicx}
\usepackage{dcolumn}
\usepackage{bm}

\newtheorem{tw}{Theorem}
\newtheorem{de}{Definition}
\newtheorem{co}{Corollary}
\newtheorem{pr}{Proposition}

\newcommand{\be}{\begin{equation}}
\newcommand{\ee}{\end{equation}}
\newcommand{\bea}{\begin{eqnarray}}
\newcommand{\eea}{\end{eqnarray}}

\begin{document}

\title[Compatible symplectic connections]{Compatible symplectic connections on a cotangent bundle  \\ and the Fedosov quantization}

\author{Jaromir Tosiek}%
\email{tosiek@p.lodz.pl}
\affiliation{Institute of Physics, Technical University of Lodz, Wolczanska 219, 90-924 Lodz, Poland.}

\date{\today}

\begin{abstract}
A global construction of a family of symplectic connections on a cotangent bundle compatible with some linear symmetric  connection on the base space  is proposed.  Examples of the compatible symplectic connections are given. A detailed analysis of an Abelian connection and of flat sections  for this kind of  symplectic connection in the Fedosov algorithm are presented. Some  properties of the  $*$-product determined by the compatible symplectic connection are shown.
\end{abstract}

\pacs{02.40.Hw, 03.65.Ca}
\keywords{symplectic connection, Fedosov deformation quantization}
\maketitle

\section{Introduction}
The first version of phase space formulation of quantum mechanics was proposed by Moyal \cite{MO49}, who adapted the ideas of Weyl \cite{WY31}, Wigner \cite{WI32} and Groenewold \cite{GW46}. Unfortunately, this 
formalism can be applied only when the phase space of a system is ${\mathbb R}^{2n}.$ The successful generalisation of Moyal's result was published by Bayen {\it et al.} \cite{baf,bay}. In these papers quantum mechanics was presented as a deformed version of classical mechanics with the Dirac constant $\hbar$ being the deformation parameter. Since then a new branch of physics and mathematics called deformation quantization has  developed rapidly.

Let $({\cal W}, \omega)$ be a symplectic manifold. The {\bf natural star product} on the manifold $({\cal W}, \omega)$ is a mapping   
 \[
 *:C^{\infty}({\cal W}) \times C^{\infty}({\cal W}) \rightarrow C^{\infty}({\cal W})[[\hbar]], \;\;\;\;\;
 (f,g)\rightarrow \sum_{i=0}^{\infty} \hbar^i B_i(f,g)
\]
 fulfilling the requirements:
\begin{enumerate}
\item
$B_i$
are bidifferential operators of order  $\leq i$ in each argument,
\item
  $B_0(f,g)=f \cdot g,$
\item
$
 B_1(f,g)=\frac{1}{2}\{f,g\}_{\rm P}$, where $\{\cdot,\cdot\}_{\rm P}$ denotes the Poisson bracket,
\item
$
 B_i(1,f)=B_i(f,1)=0.$
\end{enumerate}
The natural $*$--product, which coincides at each order with the Moyal $*$--product, is the so-called  Vey star product.

 In his outstanding works  \cite{6, 7} Boris Fedosov presented  a recurrent algorithm of construction of a  natural Vey $*$-product   
 on any symplectic manifold. The $*$-product of functions obtained in this formalism is determined by a symplectic connection on the symplectic manifold. Every symplectic manifold can be equipped with   many symplectic connections. 
The natural question arises which one  of these symplectic connections is preferable. There are several criteria for the choice (see a review \cite{biel}). One of them is the simplicity of Fedosov's recurrent formulas. Imagine that   in the $3$rd step of the recurrence on the $6$--D symplectic manifold the number of elements representing  the Abelian connection exceeds 1500! Hence the most desirable are symplectic connections generating finite Abelian connections. Unfortunately, 
 in many cases such a symplectic connection does not exist or  has a very special form \cite{ja5, ja6}. Another criterion is compatibility of the symplectic connection with some extra structures present at the symplectic manifold. For example on pseudo-K\"{a}hler manifolds the preferred symplectic connection preserves  the almost complex structure. It is also possible to select symplectic connections using a variational principle \cite{bour}.

In the  current paper we consider  symplectic manifolds which are cotangent bundles over paracompact manifolds. Symplectic manifolds of this type appear in physics as phase spaces of free systems or  systems with  scleronomic constraints. 
Our aim is to equip them with  symplectic connections which generate  relatively simple Abelian connections and, moreover, are natural from a physical point of view. 
Presented considerations are based on a conviction that Physics is determined by the 	geometry of a configuration space. 

We emphasise the fact that the problem of quantization on a cotangent bundle has  already been solved \cite{gut5, bor}. Hence our aim is not to rediscover a star product on a cotangent space but  to introduce some physically reasonable class of $*$--products on it. The fact that these products will be equivalent does not diminish their role. Equivalent $*$--products usually lead to different eigenvalues of observables, different time evolution 
of states,
and different average values of functions. Thus mathematically equivalent star products are physically distinguishable.

We show that there is a whole family of natural $*$--products satisfying our  conditions. This fact is in agreement with the observation \cite{miel} that  quantum mechanics cannot probably be reduced to a unique deformation of classical physics. 
\vspace{0.5cm}  
 
The first part of the article is  dedicated to
 the special class of  symplectic connections mentioned above. The construction  of these connections  has been done in a proper Darboux atlas, which  is the set of physically preferable charts. Yet we emphasize that this construction is global. Its result can be transformed easily into an arbitrary atlas. 
 We show that in the proper Darboux coordinates we can choose the  symplectic connection  in such a way that its coefficients with two or three momenta indices vanish. Moreover, the symplectic connection components with one momentum index are determined by the linear symmetric connection on the configuration space of the system. This is the reason why  we call this kind of symplectic  connection  the compatible symplectic connection. 
Finally, the symplectic connection coefficients with three spatial indices are homogeneous functions of momenta. 
Besides the homogeneity in the momenta we do not impose any extra conditions on these coefficients. Therefore the same linear connection on the base space leads to the whole range of
  compatible symplectic connections.

It is worth mentioning that the compatible symplectic connections have interesting geometric properties.
Let $X$ be a vector field transported parallel with respect to a compatible symplectic connection $\gamma$ 
along a curve $c$ on a cotangent bundle ${\cal T^*M}$. 
Then the canonical projection of the field $X$ 
on the base space ${\cal M}$ is parallel propagated along the projected curve $\pi(c)$
with respect to the linear connection $\Gamma$,  which the symplectic connection $\gamma $ is compatible with. 
 Coordinate- free characterization of compatible symplectic connections is presented in Theorem \ref{twfree}. 

A special case of our compatible symplectic connection is 
the induced symplectic connection $\nabla^0 $ introduced by Bordemann {\it et al} \cite{bor} and independently by 
J. F. Pleba\'{n}ski {\it et al} \cite{pleb1}. We analyse this example  in detail.

The second part of the paper is of a strictly technical character and is focused on 
the Fedosov $*$-product based on   compatible symplectic connections.
We obtain several properties of the Abelian connection generated by the compatible symplectic connections. We point out that in this case the recurrence relation leads to a relatively simple form of the Abelian connection series. 
There are only three types of elements appearing in the Abelian series $r.$ The recurrence is completely determined by the curvature $2$--form $R_{\gamma + r}$  components of the connection $\gamma + r$ standing at the elements $(y^1)^{i_1} \cdots (y^n)^{i_n} dq^{\alpha} \wedge dp^{\beta}.$ We propose some simplified algorithm of calculating the Abelian connection for  compatible symplectic connections.

Next we present a variety of  properties of the flat sections of the Weyl bundle. The flat sections are determined  by functions from the cotangent bundle ${\cal T^*M}$ and are applied  to define the $*$--product of these functions. 
In the same section
 we consider the $*$-product generated by the compatible symplectic connection of some special classes of functions. Among other things we compute the $*$-product of functions dependent only on spatial coordinates and conclude that this product reduces to  the `usual' pointwise product of them. We also   consider the Moyal brackets of positions and momenta. The commutation rules for positions and momenta are analogous to  the ones following from the Dirac quantization scheme and they are invariant under proper Darboux transformations.

If it is possible, we use the Einstein summation convention. In situations in which the nature of coordinates ( spatial or momenta ) is not important we denote them  by lower case Latin letters. The small Greek letters: $\alpha, \beta, \ldots$ represent spatial coordinates. The capital Latin letters $I,J, \ldots$ correspond to momenta coordinates. 


\section{ A symplectic connection on a cotangent bundle compatible with some linear connection on a base space}

Let $({\cal W}, \omega)$ be a $2n$--D symplectic manifold and let ${\cal A}= \{({\cal U}_z,\phi_z)\}_{z \in J}$  be an atlas on it. 
\begin{de}
 The {\bf symplectic connection} $\gamma$ on ${\cal W}$ is a torsionfree connection    satisfying  the conditions
\be
\label{d00}
\omega_{ij;k}=0,\;\;\; 1 \leq i,j,k \leq 2n, 
\ee
where the semicolon `$\: ;$' stands for the covariant derivative.
\end{de}
For  any point ${\tt p} \in {\cal W} $ there exist some local coordinates $x^1, \ldots, x^{2n}$ on a neighborhood of  ${\tt p}$ such that 
\be
\label{1}
\omega= \sum_{i=1}^{n} dx^i \wedge dx^{i+n}.
\ee
The chart $({\cal U}_z, \phi_z)$ with coordinates $x^1, \ldots, x^{2n}$ is called the {\bf Darboux chart}.
In  the Darboux coordinates the system of equations (\ref{d00}) reads
\be
\label{d1.1}
\omega_{ij;k}= - \gamma^l_{ik}\omega_{lj} - \gamma^l_{jk}\omega_{il}= \gamma_{jik}- \gamma_{ijk}=0,
\ee
where
$\gamma_{ijk} :=  \gamma^l_{jk}\omega_{il}.
$
Coefficients $\gamma_{ijk}$ are symmetric with respect to the indices $\{i,j,k\}.$ 
\begin{de}
The symplectic manifold $({\cal W}, \omega)$ endowed with the symplectic connection $\gamma$ is called a  {\bf Fedosov manifold} and it is denoted by $({\cal W}, \omega, \gamma).$
\end{de}

 A symplectic connection exists on any symplectic manifold. Moreover, every symplectic manifold may be equipped with many symplectic connections.  In a chart $({\cal U}_z,\phi_z)$ the difference
\[
\Delta_{ijk} := \gamma_{ijk} - \overline{\gamma}_{ijk}
\]
 between the coefficients $\gamma_{ijk}$ and $\overline{\gamma}_{ijk}$ of two symplectic connections $\gamma$ and $\overline{\gamma}$ on $({\cal W}, \omega)$ is a tensor of the type $(0,3)$, which is   symmetric in its indices. Hence, starting from the same symplectic manifold $({\cal W}, \omega)$  we may construct many Fedosov manifolds.

Hereafter we will work in Darboux coordinates so locally every symplectic connection $\gamma$ will be characterized by 
 the coefficients $\gamma_{ijk}$, which are  symmetric in their indices.
 The coefficients  
 $\gamma^i_{jk}=\omega^{il}\gamma_{ljk},$ 
where
\[
\omega_{kl}\omega^{lj}=\delta^j_k.
\]
The 
relations:
\[
\gamma^{\:i + n}_{j + n \:k+n}= - \,\gamma^{\:j}_{i\: k+n}= - \,\gamma^{\:k}_{i\: j+n} \;\;, \;\;
\gamma^{\:i + n}_{k \: j + n}= \gamma^{\:k + n}_{i \: j + n}= - \,\gamma^{\:j}_{i \:k}, 
\]
\be
\label{jednak}
\gamma^{\:i + n}_{j  \:k} = \gamma^{\:j + n}_{i \:k}= \gamma^{\:k + n}_{i  \:j}
\;\;, \;\; 
\gamma^{\:i }_{j+n  \:k+n} = \gamma^{\:j }_{i+n \:k+n}= \gamma^{\:k }_{i+n  \:j+n} \;\;,
\;\; 1 \leq i,j,k \leq n
\ee
are
the straightforward consequence of the symmetry of coefficients $\gamma_{ijk}$ in their indices.

Locally the symplectic curvature tensor components are defined as
\be
\label{e1}
K_{ijkl}:=\omega_{iu}K^u_{jkl}=\frac{\partial \gamma_{ijl}}{\partial x^k}
- \frac{\partial \gamma_{ijk}}{\partial x^l}+ 
\omega^{st}\gamma_{til}\gamma_{sjk}
- \omega^{st}\gamma_{tik}\gamma_{sjl}.
\ee
The coefficients $K_{ijkl}$ satisfy the  equalities \cite{gel}
\be
\label{e2}
K_{ijkl}=- K_{ijlk}\;\;\; , \;\;\; K_{ijkl}=K_{jikl},
\ee
\be
\label{e3}
K_{ijkl}+K_{iljk}+K_{iklj}=0,
\ee
\be
\label{e4}
K_{ijkl}+K_{lijk}+K_{klij}+K_{jkli}=0.
\ee
Observe that  property (\ref{e4}) follows from (\ref{e2}) and (\ref{e3}).

The second Bianchi identity  reads
\be
\label{e4.1}
K_{ijkl;m}+K_{ijmk;l}+K_{ijlm;k}=0,
\ee
where all covariant derivatives are calculated with respect to the symplectic connection.

Let our symplectic manifold be  the cotangent bundle ${\cal T^*M}$ over some configuration space ${\cal M}$. Hereafter we assume that ${\cal M}$ is an $n$--D paracompact smooth differentiable manifold.

Moreover, let $({\cal U}_z,\phi_z)$ be a local chart on the base space  ${\cal M}$. The local coordinates 
of a  point ${\tt p} \in {\cal U}_z$  
in the chart  $({\cal U}_z,\phi_z)$ are $(q^1, \ldots, q^n)$, where $n=\dim {\cal M}.$ There exists the bundle projection 
$\pi:{\cal T^*M} \rightarrow {\cal M} $ so we can introduce  coordinates on $\pi^{-1}({\cal U}_z)$ in a natural way. Indeed, let ${\bf P}=p_i dq^i$ be a cotangent vector at the point ${\tt p}.$ The coordinates $(\tilde{q}^1, \ldots,\tilde{q}^{2n}) $ of the point $({\tt p}, {\bf P}) \in \pi^{-1}({\cal U}_z)$ are defined as:
\[
\tilde{q^i}=q^i, \;\; i=1, \ldots, n, \;\;\; \tilde{q^i}= p_{i-n}, \;\; i=n+1, \ldots, 2n.
\]
The coordinates $(q^i,p_i)$ in the chart $\Big(\pi^{-1}({\cal U}_z),(q^i,p_i) \Big)$ are known as the {\bf coordinates induced by  the chart} $({\cal U}_z,\phi_z).$ In these coordinates the symplectic form determined by the basic $1$--form $\theta= p_i dq^i$ reads
\be
\label{2}
\omega := -d \theta = dq^i \wedge dp_i
\ee
so the induced coordinates $(q^i,p_i)$ are also the Darboux coordinates.  
 \begin{de}
Let $\{({\cal V}_{v}, \phi_{v})\}_{v \in J}$ be an atlas on the symplectic manifold ${\cal T^*M}$ such that in every chart the coordinates $\tilde{q}^i=q^i, \; 1 \leq i \leq n$ determine points on the base manifold ${\cal M}$ and $\tilde{q}^{i+n}=p_i, \; 1 \leq i \leq n$ denote momenta in natural coordinates. Every  atlas of this form is called a {\bf proper Darboux atlas} and every chart of this atlas is known as a   {\bf proper Darboux chart}. The transition functions define the point transformations
\be
\label{240709}
Q^k=Q^k(q^i), \;\; P_i= \frac{\partial q^l}{\partial Q^i}p_l,\;\; i,k=1, \ldots,n.
\ee
\end{de}
Assume that the base manifold ${\cal M}$ of the 
cotangent bundle ${\cal T^*M}$ is equipped with some linear connection $\Gamma$. We propose  some  global construction 
of a symplectic connection on ${\cal T^*M}$ compatible with the linear connection from the base space ${\cal M}.$ {\bf This construction is not unique}. We cover the symplectic manifold ${\cal T^*M}$ with a proper Darboux atlas. We require that in every proper Darboux atlas our symplectic connection $\gamma$ fulfills two natural conditions: 
\begin{enumerate}
\item
it `contains' 
the linear  symmetric  connection $\Gamma$
from the base space ${\cal M}$ and 
 \item
all of the symplectic connection coefficients  take  the form as simple  as possible.
\end{enumerate}
In any proper Darboux chart the first requirement means that 
$ \forall_{\alpha, \beta, \delta}\;\;\;
 \gamma^{\alpha}_{\beta \delta}= 
\Gamma^{\alpha }_{\beta \delta}.$

At the beginning we analyse the transformation rule for a symplectic connection under proper Darboux transformations.
As it is known, the general transformation rule for  symplectic connection coefficients is of the form
\be
\label{nowa1}
\gamma'_{ijk}(\tilde{Q}^1, \ldots, \tilde{Q}^{2n})=\frac{\partial \tilde{q}^l}{\partial \tilde{Q}^i}\frac{\partial \tilde{q}^r}{\partial \tilde{Q}^j}
\frac{\partial \tilde{q}^s}{\partial \tilde{Q}^k} \gamma_{lrs}(\tilde{q}^1, \ldots, \tilde{q}^{2n}) + \omega_{rd}\frac{\partial \tilde{q}^r}{\partial \tilde{Q}^i}
\frac{\partial^2 \tilde{q}^d}{\partial \tilde{Q}^j \partial \tilde{Q}^k}.
\ee
In Darboux coordinates the symplectic connection coefficients are symmetric with respect to all indices so we need to consider only four kinds of elements: $\gamma_{IJK}, \gamma_{IJ \alpha}, \gamma_{I \alpha \beta} $ and $\gamma_{\alpha \beta \delta}.$ 

Components $\gamma_{IJK}$ transform like tensors under proper Darboux transformations and are functions only of the connection coefficients of the same type. Hence if we assume that all coefficients $\gamma_{IJK}$ vanish in some proper Darboux atlas, this property remains true in any  proper Darboux atlas and it is compatible with the transformation rule.

Elements $ \gamma_{IJ \alpha}$ also transform like tensors under proper Darboux transformations. They are functions of coefficients of the same kind and  $\gamma_{IJK}.$ Thus if all terms $\gamma_{IJK}=0$ as we assumed before and all
$ \gamma_{IJ \alpha}=0$ in some proper Darboux atlas, the choice of the connection coefficients is consistent.

Let us consider the coefficients $ \gamma_{I \alpha \beta}.$ In general they depend on coefficients of the same kind, $\gamma_{IJK}$ and 
$ \gamma_{IJ \alpha}.$ Remembering previous results we see that this relation reduces to the dependency on elements of the same type. The nontensorial part   $\omega_{rd}\frac{\partial q^r}{\partial Q^i}
\frac{\partial^2 q^d}{\partial Q^j \partial Q^k}$ of the transformation formula does not vanish. It is a function only of spatial coordinates. These two facts suggest the following definition of $ \gamma_{I \alpha \beta}.$ Let $\Gamma^{\delta}_{\alpha \beta}$  be   linear symmetric connection coefficients on the base manifold ${\cal M}.$ Then in proper Darboux coordinates we define
\be
\label{nowa2}
\gamma_{I \alpha \beta}(q^1, \ldots,q^n,p_1, \ldots, p_n) :=- \Gamma^{I-n}_{\alpha \beta}(q^1, \ldots,q^n).
\ee
This definition establishes the relation between the linear symmetric connection from the configuration space ${\cal M}$ and the symplectic connection on the phase space ${\cal T^*M}.$ The symplectic connection coefficients $\gamma_{I \alpha \beta}$ are determined by the linear symmetric connection  on ${\cal M}$ and depend only on the spatial coordinates $q^1, \ldots,q^n.$

Finally we consider the symplectic connection coefficients $\gamma_{\alpha \beta \delta}.$ 
From (\ref{nowa1}) we see that they depend on the terms $\gamma_{I \alpha \beta}$ and $\gamma_{\alpha \beta \delta}.$ Moreover, the transformation rule applied to coefficients $\gamma_{I \alpha \beta}$ leads to elements of the form $P_{\alpha}f^{\alpha}(Q^1, \ldots, Q^n).$ The consequence of the transformation rule (\ref{nowa1}) is also the presence of the expression
$P_{\alpha}g^{\alpha}(Q^1, \ldots, Q^n)$ following from the nontensorial part of this rule. Thus it is expected that in every proper Darboux chart the symplectic connection coefficients $\gamma_{\alpha \beta \delta}$ should be  of the form
\be
\label{nowa3}
\gamma_{\alpha \beta \delta}(q^1, \ldots,q^n,p_1, \ldots, p_n)= p_{\varepsilon }f^{\varepsilon }_{\alpha \beta \delta}(q^1, \ldots, q^n).
\ee 
Quantities $f^{\varepsilon }_{\alpha \beta \delta}(q^1, \ldots, q^n)$ are purely differentiable geometrical objects, which are symmetric with respect to the indices $\{\alpha, \beta, \delta\}.$ Under proper Darboux transformations they obey the following transformation rule
\[
{f'}^{\varepsilon }_{\alpha \beta \delta}(Q^1, \ldots, Q^n)=\frac{\partial q^{\eta}}{\partial Q^{\alpha}}
\frac{\partial q^{\tau}}{\partial Q^{\beta}} \frac{\partial q^{\upsilon}}{\partial Q^{\delta}}
\frac{\partial Q^{\varepsilon}}{\partial q^{\mu }}\: f^{\mu}_{\eta \tau \upsilon}(q^1, \ldots, q^n)+
\]
\[
+ 
\sum_{\kappa=1}^{n}
\frac{\partial^2 Q^{\varepsilon}}{\partial q^{\mu} \partial q^{\kappa }}\left( 
\frac{\partial q^{\mu}}{\partial Q^{\alpha}}
\frac{\partial q^{\tau}}{\partial Q^{\beta}} \frac{\partial q^{\upsilon}}{\partial Q^{\delta}}+
\frac{\partial q^{\tau}}{\partial Q^{\alpha}}
\frac{\partial q^{\upsilon}}{\partial Q^{\beta}} \frac{\partial q^{\mu}}{\partial Q^{\delta}}+
\frac{\partial q^{\tau}}{\partial Q^{\alpha}}
\frac{\partial q^{\mu}}{\partial Q^{\beta}} \frac{\partial q^{\upsilon}}{\partial Q^{\delta}}
\right)\gamma_{\kappa + n \: \tau \upsilon}(q^1, \ldots, q^n)+
\]
\be
+ \frac{\partial q^{\eta}}{\partial Q^{\alpha}}
\frac{\partial q^{\tau}}{\partial Q^{\beta}} \frac{\partial q^{\upsilon}}{\partial Q^{\delta}}
\frac{\partial^3 Q^{\varepsilon}}{\partial q^{\eta} \partial q^{\tau} \partial q^{\upsilon}}.
\ee

The last step in our construction is to put together the coefficients $\gamma_{\alpha \beta \delta}$ on intersections of charts. Since the base manifold is paracompact we can use the partition of unity. Let $\{g_z \}_{z \in J}$ be a partition of unity corresponding to an atlas $\{({\cal U}_{z}, \phi_{z})\}_{z \in J}$ on ${\cal M}.$ Then we define
\[
\gamma_{\alpha \beta \delta}:=\sum_{z \in J} g_z (\gamma_{\alpha \beta \delta})_z.
\]
Locally these coefficients are of the form (\ref{nowa3}). Notice that they disappear on the base space ${\cal M}.$

The construction of the symplectic connection $\gamma$ on the cotangent bundle ${\cal T^*M}$   has been completed. 
This natural symplectic connection will be called   the {\bf compatible symplectic connection}.
Although our considerations have been  made locally, the proposed construction of the compatible symplectic connection is global (compare the paragraph about existence and extension of connections  \cite{koba}).

A symplectic connection on a $2$--D symplectic manifold in a Darboux chart is determined by  $\frac{2}{3}n(n+1)(2n+1) $ coefficients. In the case  of a compatible symplectic
at least a half of them vanishes.

The compatible symplectic connection  coefficients  $\gamma^i_{jk}$ on the cotangent bundle ${\cal T^*M}$ in a proper Darboux chart are:
\[
\gamma^{\alpha}_{JK}= \gamma^{I}_{JK}=\gamma^{\alpha}_{J\beta}=0,
\]
\[
\gamma^{\beta + n}_{I \alpha}=
\gamma^{\alpha + n}_{I \beta}=- \gamma^{I-n}_{\alpha \beta}= \gamma_{I \alpha \beta}=-\Gamma^{I-n}_{\alpha \beta},
\]
\be
\label{nowa3.5}
\gamma^{\beta + n}_{ \alpha \delta}=\gamma^{\delta + n}_{ \alpha \beta }=
 \gamma^{\alpha + n}_{ \beta \delta}= \gamma_{\alpha \beta \delta}. 
\ee
This result is in agreement with (\ref{jednak}).

The compatible symplectic connections have  nice geometric properties. Let $c:\tilde{q}^i=\tilde{q}^i(t), \; i=1, \ldots, 2n, \; t \in {\mathbb R}$
be a smooth curve on the cotangent bundle ${\cal T^*M}$. Moreover, let $X$  denote  a vector field parallel  along this curve.
Locally it means that 
\be
\label{czw1}
\frac{d X^i}{dt}+\gamma^i_{jk} X^j \frac{d \tilde{q}^k}{dt}=0.
\ee
Then using  relations (\ref{nowa3.5}) 
we see that the set of equations (\ref{czw1}) can be divided in two parts: 
the conditions 
\be
\label{czw2}
\frac{d X^{\alpha}}{dt}+ \gamma^{\alpha}_{ij} X^{i}\frac{d \tilde{q}^{j}}{dt}\stackrel{\rm (\ref{nowa3.5})}{=}
\frac{d X^{\alpha}}{dt}+ \Gamma^{\alpha}_{\beta \delta} X^{\beta}\frac{d q^{\delta}}{dt}=0
\ee
 and
some other relations of the kind
\[
\frac{d X^I}{dt}+\gamma^I_{jk} X^j\frac{d \tilde{q}^k}{dt}=0.
\]
Now we project the curve $c$ in the canonical way on the base space ${\cal M}.$ This projected curve $\pi(c)$ is described by the system of equations $\pi(c):q^{\alpha} =q^{\alpha}(t), \; \alpha=1, \ldots, n, \; t \in {\mathbb R}.$ The vector field $X$ transforms under this canonical projection into a vector field $\pi_*(X).$ As we can see from (\ref{czw2}) the vector field $\pi_*(X)$ is parallel propagated on the curve $\pi(c).$ Thus we conclude that 
\begin{pr}
\label{prop1}
Every compatible symplectic connection guarantees that for every vector field $X$  parallel transported along  the curve $c$ on the cotangent bundle ${\cal T^*M}$ the canonical projection $\pi_*(X)$ of this field  on the base space ${\cal M}$ is  parallel propagated along  the projected  curve $\pi(c).$
\end{pr}

Moreover, if we consider the vector field $X^i= \frac{d\tilde{q}^i(t)}{dt},$  we see that
the canonical projection of any geodesic in ${\cal T^*M}$ with respect to the compatible symplectic connection $ \gamma$ onto the base space ${\cal M}$ is the geodesic with respect to the linear symmetric connection $\Gamma$ with the same affine parameter.

Let $\theta $ denote, as before, the basic $1-$form on the cotangent bundle  ${\cal T^*M}.$ The canonical vector field ${ V}$ on ${\cal T^*M}$ is defined by the relation
\[
i_{ V}\omega = - \theta.
\]
Locally in Darboux coordinates
${ V}= p_{\epsilon} \frac{\partial }{\partial p_{\epsilon}}.$
\begin{de}
If for all vector fields
${ X}, { Y} \in \Gamma(T^1_0( {\cal T^*M}))$ the following equality holds
\[
{\cal L}_{ V}\nabla_{ X} { Y}-\nabla_{{\cal L}_{ V}{ X}}{ Y}-\nabla_{ X}
{\cal L}_{ V}{ Y}=0,   
\]
then the connection $\nabla $ on the cotangent bundle  ${\cal T^*M}$ is  {\bf homogeneous}.
\newline
The symbol ${\cal L}_{ V}$ stands for the Lie derivative with respect to the canonical vector field $V.$
\end{de}
In a proper Darboux chart it means that $ \forall_{\alpha,I} $
\[
p_{\epsilon}\frac{\partial \Gamma^{\alpha}_{ \beta \delta}}{\partial p_{\epsilon}}X^{\beta} Y^{\delta}+
p_{\epsilon}\frac{\partial \Gamma^{\alpha}_{ J \beta}}{\partial p_{\epsilon}}X^{J} Y^{\beta}+
p_{\epsilon}\frac{\partial \Gamma^{\alpha}_{ \beta J}}{\partial p_{\epsilon}}X^{\beta} Y^{J}+
p_{\epsilon}\frac{\partial \Gamma^{\alpha}_{ JL}}{\partial p_{\epsilon}}X^{J} Y^{L}+
\]
\be
\label{niedz1}
+2  \Gamma_{JL}^{\alpha}X^J Y^L +  \Gamma_{J\beta}^{\alpha}X^J Y^{\beta}+  \Gamma_{\beta J}^{\alpha}X^{\beta} Y^J=0
\ee
and
\be
\label{niedz2}
p_{\epsilon}\frac{\partial \Gamma^{I}_{ \beta \delta }}{\partial p_{\epsilon}}X^{\beta}Y^{\delta}
+ p_{\epsilon}\frac{\partial \Gamma^{I}_{J \beta  }}{\partial p_{\epsilon}}X^{J}Y^{\beta}+
p_{\epsilon}\frac{\partial \Gamma^{I}_{\beta J }}{\partial p_{\epsilon}}X^{\beta}Y^{J} +
p_{\epsilon}\frac{\partial \Gamma^{I}_{JL }}{\partial p_{\epsilon}}X^{J}Y^{L}
- \Gamma^{I}_{ \beta \delta } X^{\beta}Y^{\delta}+ 
\Gamma^{I}_{JL } X^{J}Y^{L}=0,
\ee
where symbols $\Gamma^i_{jk}$ denote coefficients of the connection $\nabla.$ 
Substituting  (\ref{nowa3.5}) into (\ref{niedz1}) and (\ref{niedz2}) we conclude that the proposition holds. 
\begin{pr}
\label{prop2}
Every compatible symplectic connection is homogeneous.
\end{pr}

Let $\tau \stackrel{\rm locally }{=} \tau_{\epsilon}dq^{\epsilon}$ be a $1$--form defined on the base space ${\cal M}.$ By a {\bf vertical lift} of the form $\tau$  we understand the vector field $\tau^V$ on the cotangent bundle ${\cal T^*M}$ 
\be
\label{61220102}
\tau^V \stackrel{\rm locally }{=} \sum_{\epsilon =1}^n \tau_{\epsilon} \frac{\partial}{\partial p_{\epsilon}}.
\ee
Geometric definition of the vertical lift has been presented  by Yano and Ishihara \cite{yano}.

Moreover, let $X \stackrel{\rm locally }{=} X^{\epsilon} \frac{\partial}{\partial q^{\epsilon}}$ denote a vector field on the base space ${\cal M}.$  A {\bf horizontal lift} of the vector field $X$ is the vector field $X^H$ on the cotangent bundle ${\cal T^*M}$ 
\be
\label{61220103}
X^H \stackrel{\rm locally }{=}  X^{\epsilon} \frac{\partial}{\partial q^{\epsilon}} + p_{\alpha} \Gamma^{\alpha}_{\beta \epsilon} X^{\epsilon}  \frac{\partial}{\partial p_{\beta}}.
\ee
By $ \Gamma^{\alpha}_{\beta \epsilon}$ we denote coefficients of a  linear symmetric connection $\Gamma$ on the base manifold ${\cal M}.$ The definition of the horizontal lift can be seen in ref. \cite{yano}.

The covariant differentiation with respect to a connection $\tilde{\Gamma}$ will be represented by the symbol $\nabla^{(\tilde{\Gamma})}.$  After simple calculations we see that the following equalities hold.
\begin{pr}
\label{prop3}
Let $\Gamma$ be a linear symmetric connection on a  manifold ${\cal M}$ and $\gamma$ be a symplectic connection on the cotangent bundle ${\cal T^* M}$ compatible with the connection $\Gamma.$ Then for every $1$-- form $\tau, \sigma$  and every vector field $X$ defined on the  manifold ${\cal M}$
\be
\label{71020101}
\nabla^{(\gamma)}_{\sigma^V} \tau^V=0 \;\;,\;\; \nabla^{(\gamma)}_{\sigma^V} X^H=0 \;\;,\;\; \nabla^{(\gamma)}_{X^H} \tau^V= \Big( \nabla^{(\Gamma)}_X \tau \Big)^V.
\ee
\end{pr}
We formulate the  theorem.
\begin{tw}
\label{twfree}
A symplectic connection $\gamma$ on a cotangent bundle ${\cal T^* M}$ is compatible with a linear symmetric connection $\Gamma$ on the base space ${\cal M}$ iff
 it is homogeneous and 
satisfies  the conditions 
\be
\label{61220101}
\nabla^{(\gamma)}_{\sigma^V} \tau^V=0 \;\;,\;\; \nabla^{(\gamma)}_{\sigma^V} X^H=0.
\ee
\end{tw}
\underline{Proof}

First, assume that
 $\gamma$ is a symplectic connection compatible with a linear symmetric connection $\Gamma$. Then, from 
Propositions  \ref{prop2} and \ref{prop3}, it is homogeneous and it satisfies conditions (\ref{61220101}).

Conversely, assume that a symplectic homogeneous connection 
$\gamma$ on ${\cal T^*M}$ fulfills conditions
 (\ref{61220101}). Write the relation $\nabla^{(\gamma)}_{\sigma^V} \tau^V=0$ in a proper Darboux chart. From (\ref{61220102}) we get that the spatial component $\beta$ of $\nabla^{(\gamma)}_{\sigma^V} \tau^V$ equals 
\[
\Big(\nabla^{(\gamma)}_{\sigma^V} \tau^V \Big)^{\beta}= \sum_{\alpha , \epsilon=1}^n\, \gamma^{\,\beta}_{\alpha+n \, \epsilon +n} \; \sigma_{\alpha} \tau_{\epsilon}= 0
\]
for arbitrary 1-- forms $\sigma, \tau$. Thus there must be $\forall \; \alpha, \beta, \epsilon \; \;\; \gamma^{\beta}_{\alpha+n \, \epsilon +n}=0.$ 

Now
\[
\Big(\nabla^{(\gamma)}_{\sigma^V} \tau^V \Big)^{\beta+n}= \sum_{\alpha , \epsilon=1}^n\, \gamma^{\,\beta+n}_{\alpha+n \, \epsilon +n} \; \sigma_{\alpha} \tau_{\epsilon}= 0.
\]
We see that $\forall \; \alpha, \beta, \epsilon \; \;\; \gamma^{\beta+n}_{\alpha+n \, \epsilon +n}=0.$ Moreover, from (\ref{jednak}) we obtain that $\forall \; \alpha, \beta, \epsilon \; \;\; \gamma^{\,\beta}_{\alpha \, \epsilon +n}=0.$

The relation 
\[
\Big( \nabla^{(\gamma)}_{\sigma^V} X^H \Big)^{\beta +n}= \sum_{\alpha, \epsilon=1}^n \Gamma^{\alpha}_{\beta \,\epsilon} \sigma_{\alpha} X^{\epsilon}+ \sum_{\alpha, \epsilon =1}^n \,
\gamma^{\beta+n}_{\alpha +n\, \epsilon} \sigma_{\alpha} X^{\epsilon}+ \sum_{\alpha, \delta, \epsilon, \tau =1}^n\, p_{\delta} \Gamma^{\delta}_{\tau \, \epsilon} \, \gamma^{\beta + n}_{\alpha + n \, \tau + n } \sigma_{\alpha} X^{\epsilon}
=0
\]
can be simplified, due to the previous result that all coefficients $\gamma^{\beta+n}_{\alpha+n \, \epsilon +n}$ disappear. Thus
\[
 \sum_{\alpha, \epsilon =1}^n \, \Big( \Gamma^{\alpha}_{\beta \,\epsilon} +
\gamma^{\beta+n}_{\alpha +n\, \epsilon} \Big) \sigma_{\alpha} X^{\epsilon}=0.
\]
Hence, $\forall \; \alpha, \beta, \epsilon \; \;\; \gamma^{\beta+n}_{\alpha +n\, \epsilon} = - \Gamma^{\alpha}_{\beta \,\epsilon}. $ Applying (\ref{jednak}) we conclude that also $  \forall \; \alpha, \beta, \epsilon \; \;\;  \Gamma^{\alpha}_{\beta \,\epsilon}= \gamma^{\alpha}_{\beta \,\epsilon}.$

Finally,  consider the fact, that our symplectic connection is homogeneous. Applying previous results we find that Eq. (\ref{niedz2}) reduces to the form
\be
\label{71220101}
p_{\epsilon} \frac{\partial \gamma^{I}_{\alpha \beta}}{\partial p_{\epsilon}}- \gamma^{I}_{\alpha \beta}=0.
\ee
A general solution of it is a function $\gamma^{I}_{\alpha \beta}=\frac{1}{\sqrt{\sum_{\epsilon=1}^n p_{\epsilon}^2}}f \Big(\frac{p_1}{\sqrt{\sum_{\epsilon=1}^n p_{\epsilon}^2}}, \ldots, \frac{p_n}{\sqrt{\sum_{\epsilon=1}^n p_{\epsilon}^2}}, q^1, \ldots, q^n \Big)$ homogeneous in momenta. We assume that  the coefficients of the symplectic connection $\gamma$ are smooth functions on ${\cal T^*M}$. Hence for the fixed indices $I, \alpha, \beta$ the function $\gamma^{I}_{\alpha \beta}$ is differentiable at every point $(p_1=0,\ldots, p_n=0, q^1, \ldots, q^n).$ From formula (\ref{71220101}) we obtain that $\gamma^{I}_{\alpha \beta}(0,\ldots, 0, q^1, \ldots, q^n)=0.$ The directional derivative of $\gamma^{I}_{\alpha \beta}$ in the direction $\vec{v}= \Big(\frac{v_1}{\sqrt{\sum_{i=1}^n v_i^2}}, \ldots,
\frac{v_n}{\sqrt{\sum_{i=1}^n v_i^2}},0, \ldots,0 \Big)$ at the point 
$(0,\ldots, 0, q^1, \ldots, q^n)$
equals
\[
\lim_{t \rightarrow 0^+ } \frac{\gamma^I_{\alpha \beta} \Big(t \cdot \frac{v_1}{\sqrt{\sum_{i=1}^n v_i^2}}, \ldots, t \cdot \frac{v_n}{\sqrt{\sum_{i=1}^n v_i^2}},q^1, \ldots, q^n \Big)-\gamma^I_{\alpha \beta}(0, \ldots, 0 , q^1, \ldots, q^n)}{t}=
\]
\be
\label{81220101}
= \gamma^I_{\alpha \beta}\left( \frac{v_1}{\sqrt{\sum_{i=1}^n v_i^2}}, \ldots,  \frac{v_n}{\sqrt{\sum_{i=1}^n v_i^2}},q^1, \ldots, q^n \right).
\ee
On the other hand the directional derivative  $(\ref{81220101})$  at the point $(0, \ldots,0, q^1, \ldots, q^n)$ 
can be written as
\be
\label{81220102}
\sum_{\epsilon=1}^n \frac{\partial \gamma^I_{\alpha \beta}(0, \ldots, 0 , q^1, \ldots, q^n)}{\partial p_{\epsilon}} \cdot \frac{v_{\epsilon}}{\sqrt{\sum_{i=1}^n v_i^2}}.
\ee 
Comparing results (\ref{81220101}) and (\ref{81220102}) we conclude that locally
\[
\gamma^I_{\alpha \beta}= \sum_{\epsilon=1}^{n} p_{\epsilon} f^{I\, \epsilon}_{\alpha \beta} (q^1, \ldots, q^n).
\]
Thus we proved that the symplectic homogeneous connection $\gamma$ which fulfills conditions (\ref{61220101}) is compatible with the linear connection $\Gamma$ on the base space ${\cal M}.$
 \rule{2mm}{2mm}

In proper Darboux coordinates 
the nonvanishing components of the symplectic curvature tensor $K$ of the symplectic connection (\ref{nowa3.5}) are
$ K_{I\:\beta \gamma \delta}, K_{\alpha\beta \gamma \: I} $ and $K_{\alpha \beta \gamma \delta}.$
The coefficients of the first group are determined by the  curvature tensor of the linear symmetric connection from the base manifold ${\cal M}.$ It is easy to show that 
\[
K_{\alpha+n\:\beta \gamma \delta} (q^1, \ldots,q^n,p_1, \ldots, p_n)=-R^{\alpha}_{\beta \gamma \delta}(q^1, \ldots,q^n).
\]
Thus we conclude that the coefficients $K_{\alpha+n\:\beta \gamma \delta}$ do not
 depend on momenta. By $R^{\alpha}_{\beta \gamma \delta}$ we mean components of the curvature tensor of the linear symmetric connection $\Gamma.$

Moreover, from (\ref{e2}) and (\ref{e3}) we see that
\be
\label{nowa4}
K_{I \:\beta \gamma \delta}= K_{\beta \delta \gamma \: I} - K_{\beta  \gamma \delta \: I}.
\ee
The components $K_{\beta \delta \gamma \: I}$ are functions  of spatial coordinates only.

Finally the elements $K_{\alpha \beta \gamma \delta}$ are of the form
$
K_{\alpha \beta \gamma \delta}= \sum_{\epsilon =1}^n p_{\epsilon} (K_{\alpha \beta \gamma \delta})_{\epsilon},
$
where $\forall_{1 \leq \epsilon \leq n} \: (K_{\alpha \beta \gamma \delta})_{\epsilon}$ are  functions of variables $ q^1, \ldots, q^n.$

From the second Bianchi identity (\ref{e4.1}) we get
\[
(K_{\alpha \beta \gamma \delta})_{\epsilon}=\frac{\partial K_{\alpha \beta \gamma \epsilon +n}}{\partial q^{\delta}}-
\frac{\partial K_{\alpha \beta \delta \epsilon +n}}{\partial q^{\gamma}} +\sum_{\upsilon =1}^n \gamma_{\epsilon + n \: \upsilon \alpha}\Big(K_{\beta \delta \gamma \upsilon +n} - K_{\beta  \gamma \delta \upsilon +n}\Big)
+\sum_{\upsilon =1}^n \gamma_{\epsilon + n \: \upsilon \beta} \Big( K_{\alpha \delta \gamma \upsilon +n} 
-K_{\alpha  \gamma \delta \upsilon +n} \Big)+
\]
\be
+ \sum_{\upsilon =1}^n \gamma_{\upsilon + n \: \alpha \delta}K_{\upsilon \beta \gamma  \epsilon +n}
+  \sum_{\upsilon =1}^n \gamma_{\upsilon + n \: \beta \delta}K_{\upsilon \alpha  \gamma  \epsilon +n}
\label{nowa5}
-\sum_{\upsilon =1}^n \gamma_{\upsilon + n \: \alpha \gamma}K_{\upsilon \beta \delta  \epsilon +n}
-\sum_{\upsilon =1}^n \gamma_{\upsilon + n \: \beta \gamma}K_{\upsilon \alpha \delta  \epsilon +n}.
\ee
Hence we conclude that to obtain the complete symplectic curvature tensor we need to know all of the curvature tensor components $K_{\alpha \beta \gamma I}$ and the linear connection on the base manifold ${\cal M}.$ Flatness of   the base space ${\cal M}$  is the necessary but not sufficient condition for the cotangent bundle ${\cal T^*M}$ to be symplectic flat.

As it can be seen from (\ref{nowa4}) and (\ref{nowa5}) the necessary and sufficient condition for a compatible symplectic connection to be flat is that
\be
\label{ponadto1}
\forall_{I, \beta, \gamma, \delta} \;\; K_{\beta \gamma \delta I}=0. 
\ee
From   property (\ref{e4}) and  Definition (\ref{e1}) of the symplectic curvature tensor components we see that
the requirement (\ref{ponadto1}) is equivalent to the following statement.
\begin{tw}
\label{ponadto2}
A symplectic connection $\gamma$ compatible with a linear symmetric connection $\Gamma$ is flat if and only if the linear symmetric connection $ \Gamma$ is flat and 
in every  proper Darboux chart the coefficients $\gamma_{\alpha \beta \delta}$ of the compatible symplectic connection
are equal 
\[
\gamma_{\alpha \beta \delta}= \frac{1}{3} \sum_{\epsilon ,\upsilon =1}^n p_{\epsilon }
\left(\frac{\partial \gamma_{\epsilon +n\:\beta \delta}}{\partial q^{\alpha }} + \frac{\partial \gamma_{\epsilon +n\:\alpha \beta }}{\partial q^{\delta}} + \frac{\partial \gamma_{\epsilon +n\:\alpha \delta}}{  \partial q^{\beta} } +
2\gamma_{\epsilon +n\:\upsilon \alpha }\gamma_{\upsilon +n\:\beta \delta}+2\gamma_{\epsilon +n\:\upsilon \delta}\gamma_{\upsilon +n\:\alpha \beta }+2\gamma_{\epsilon +n\:\upsilon \beta }\gamma_{\upsilon +n\:\alpha \delta}
\right).
\]
\end{tw}
This result will also appear in the next section. 

Moreover, from  formula (\ref{e1}) we see, that  two  symplectic connections $\gamma$ and $\tilde{\gamma}$  compatible with the same linear symmetric connection $\Gamma$ but having  different components $\gamma_{\alpha \beta \delta},$ are characterised by different  symplectic curvature tensors.

The  symplectic Ricci tensor $K_{ij} := \omega^{ls}K_{lisj}.$ In a cotangent bundle equipped  with a compatible symplectic connection  only   components 
\be
\label{q0}
K_{\alpha \beta}= - \sum_{\epsilon=1}^n K_{\alpha  \beta \epsilon \: \epsilon +n}
\ee 
 can be different from $0.$
Thus the Fedosov manifold $({\cal T^*M}, \omega, \gamma)$  with a compatible symplectic connection $\gamma$ is Ricci flat if and only if  in a proper Darboux atlas 
$
\forall_{\alpha, \beta}\;\; \sum_{\epsilon=1}^n K_{\alpha  \beta \epsilon \: \epsilon +n}=0.
$

\section{Examples of compatible symplectic connections}

In the first part of this section we consider a
  symplectic connection induced by some linear connection. This kind of  symplectic connection is a special case of the compatible symplectic connection.
  The example is based on ideas presented by Yano and Ishihara \cite{yano}. A similar problem was considered by Pleba\'nski {\it et al} \cite{pleb1}. It also appeared in another context in article \cite{bor}. 

Assume that the $n$--D base manifold ${\cal M}$ is endowed with a linear symmetric connection $\Gamma.$ The tensor field $\tilde{g} \in \Gamma(T^0_2({\cal T^*M}))$ in proper Darboux coordinates is defined  as
\be
\label{n1}
\tilde{g}_{jk}=
\left(
\begin{array}{cc}
-2 p_{\epsilon} \Gamma^{\epsilon}_{\alpha \beta} & {\bf 1} \\
{\bf 1}  & {\bf 0}
\end{array}
\right).
\ee
By ${\bf 1}$ and ${\bf 0}$ we denote the identity and zero matrices of the dimension $n \times n$ respectively. The coefficients  
$\Gamma^{\epsilon}_{\alpha \beta}$ are components of the linear symmetric connection  on ${\cal M}.$ 

The tensor field $\tilde{g}$ of the type $(0,2)$ is symmetric and nondegenerate. Its signature is $(\underbrace{+, \ldots, +}_{n\; {\rm times}}, \underbrace{-, \ldots, -}_{n\; {\rm times}}).$
 The pair $({\cal T^*M},\tilde{g})$ is a $2n$--D Riemannian manifold. The Levi-- Civita connection $\tilde{\Gamma}$ on it is determined by the tensor $\tilde{g}$ according to the well known relation
\be
\label{wz1}
\tilde{\Gamma}^i_{jk}=\frac{1}{2}\tilde{g}^{il}\left(\frac{\partial \tilde{g}_{lk}}{\partial \tilde{q}^j} +
\frac{\partial \tilde{g}_{lj}}{\partial \tilde{q}^k}-
\frac{\partial \tilde{g}_{jk}}{\partial \tilde{q}^l}
\right).
\ee
By $\tilde{g}^{il}$ we mean components of the tensor inverse to the metric tensor $\tilde{g}.$

It can be easily checked  that the coefficients of the Levi-- Civita connection on the manifold $({\cal T^*M},\tilde{g})$ are
\[
\tilde{\Gamma}^{\alpha}_{\beta \delta}=\Gamma^{\alpha}_{\beta \delta}\;\;, \;\; \tilde{\Gamma}^{\alpha}_{\beta I}=0 \;\;, \;\; \tilde{\Gamma}^{\alpha}_{IJ}=0, 
\]
\be
\tilde{\Gamma}^{\alpha +n}_{\beta \delta}=p_{\epsilon} \left(\frac{\partial \Gamma^{\epsilon}_{\beta \delta}}{\partial q^{\alpha}}-\frac{\partial \Gamma^{\epsilon}_{\alpha \beta}}{\partial q^{\delta}}-\frac{\partial\Gamma^{\epsilon}_{\delta \alpha}}{\partial q^{\beta}}+2 \Gamma^{\epsilon}_{\upsilon \alpha}\Gamma^{\upsilon}_{\beta \delta}\right)\;\;,\;\;
\label{3}
\tilde{\Gamma}^{\alpha+n}_{\beta \:\delta +n}=-\Gamma^{\delta}_{\alpha \beta} \;\;, \;\; \tilde{\Gamma}^{I}_{JK}=0.
\ee

Let us lower the upper index of Christoffel symbols $\tilde{\Gamma}^{\,i}_{jk}$ by their contraction with the symplectic form. The coefficient $\tilde{\Gamma}_{ijk}:=\omega_{il}\tilde{\Gamma}^{\,l}_{jk}.$
\[
\tilde{\Gamma}_{\alpha +n\: \beta \delta}=-\Gamma^{\alpha}_{\beta \delta}\;\;, \;\;  
\tilde{\Gamma}_{I J \alpha}=0 \;\;,  \;\;
\tilde{\Gamma}_{IJK}=0, 
\]
\[
\tilde{\Gamma}_{\alpha \beta \delta}
=p_{\epsilon} \left(\frac{\partial \Gamma^{\epsilon}_{ \beta \delta}}{\partial q^{\alpha}}-\frac{\partial\Gamma^{\epsilon}_{\alpha \beta}}{\partial q^{\delta}}-\frac{\partial \Gamma^{\epsilon}_{\delta \alpha}}{\partial q^{\beta}}+2 \Gamma^{\epsilon}_{\upsilon \alpha}\Gamma^{\upsilon}_{\beta \delta}\right)\;\;,
\;\;
\tilde{\Gamma}_{\alpha \beta \:\delta+n}= 
-\Gamma^{\delta}_{ \alpha \beta} \;\;,\;\;
\tilde{\Gamma}_{\alpha IJ}=0.
\]
It is known \cite{gel} that  a symplectic manifold $({\cal W}, \omega)$ endowed with some symmetric affine connection $\Gamma$  can be equipped with a symplectic connection.
In any Darboux chart the coefficients of that symplectic connection are equal
\be
\label{e5}
\gamma_{ijk}:=\frac{1}{3}\left( \Gamma_{ijk} + \Gamma_{jik}+ \Gamma_{kij} \right).
\ee
Applying the formula (\ref{e5}) to the Levi-- Civita connection $\tilde{\Gamma}$ on the manifold $({\cal T^*M},\tilde{g})$ we obtain
the symplectic connection on ${\cal T^*M}$ induced by the Levi-- Civita connection. Moreover, since the metric tensor $\tilde{g}$  (\ref{n1}) is a function of the linear symmetric connection $\Gamma$ on the  configuration space 
${\cal M}$ then in fact the symplectic connection $\gamma$ is determined by the connection on the base space 
${\cal M}.$

In  proper Darboux coordinates the coefficients of the induced symplectic connection on ${\cal T^*M}$ read
\[
\gamma_{\,\alpha +n\:\beta \delta}= - \Gamma^{\alpha}_{\beta \delta}, \;\;\; \gamma_{\,IJ \alpha}=0,\;\; \gamma_{\,IJK}=0,
\]
\be
\label{wzor1}
\gamma_{\alpha \beta \delta}= -\frac{1}{3} p_{\epsilon} \left(\frac{\partial \Gamma^{\epsilon}_{\beta \delta}}{\partial q^{\alpha}} + \frac{\partial \Gamma^{\epsilon}_{\alpha  \beta}}{\partial q^{\delta}} + \frac{\partial \Gamma^{\epsilon}_{\alpha \delta}}{  \partial q^{\beta}} -
2\Gamma^{\epsilon}_{\upsilon \alpha}\Gamma^{\upsilon}_{\beta \delta}-2\Gamma^{\epsilon}_{\upsilon \delta}\Gamma^{\upsilon}_{ \alpha \beta}-2\Gamma^{\epsilon}_{\upsilon \beta}\Gamma^{\upsilon}_{\alpha \delta}
\right).
\ee
The induced symplectic connection $\gamma$ and the Levi-- Civita connection $\tilde{\Gamma}$ on ${\cal T^*M}$ are different. 

The nonvanishing components of the symplectic curvature tensor $K$ of the symplectic connection (\ref{wzor1}) in proper Darboux coordinates are
\[
K_{\alpha+n\:\beta \gamma \delta}=-R^{\alpha}_{\beta \gamma \delta}\;\;, \;\; K_{\alpha \beta \gamma \:\delta+n}= \frac{1}{3}\Big(R^{\delta}_{\alpha \beta \gamma }+R^{\delta}_{\beta  \alpha \gamma } \Big),
\]
\be
\label{e6}
K_{\alpha \beta \gamma \delta}= - \frac{1}{3}p_{\epsilon}\Big(R^{\epsilon}_{\beta \gamma \delta;\alpha} +R^{\epsilon}_{\alpha \gamma \delta;\beta } +3\Gamma^{\epsilon}_{\upsilon \alpha}R^{\upsilon}_{\beta \gamma \delta}+ 3\Gamma^{\epsilon}_{\upsilon \beta }R^{\upsilon}_{\alpha \gamma \delta} 
+(R^{\upsilon}_{\alpha \beta \gamma }+ R^{\upsilon}_{\beta  \alpha \gamma })\Gamma^{\epsilon}_{\delta \upsilon}- (R^{\upsilon}_{\alpha \beta \delta}+ R^{\upsilon}_{\beta  \alpha \delta})\Gamma^{\epsilon}_{\gamma \upsilon}
\Big).
\ee
By $R^{\alpha}_{\beta \gamma \delta}$ we understand components of the Riemannian curvature tensor of the connection $\tilde{\Gamma}.$

From  formulas (\ref{wzor1}) and (\ref{e6}) we see that, unlike the general case of a compatible symplectic connection, in the considered example the symplectic connection and its curvature are determined by components $\gamma_{\alpha +n\:\beta \delta },\; K_{\alpha +n\:\beta  \gamma \delta}$ and their derivatives only. Indeed,
\be
\label{e7}
\gamma_{\alpha \beta \delta}= \frac{1}{3} \sum_{\epsilon ,\upsilon =1}^n p_{\epsilon }
\left(\frac{\partial \gamma_{\epsilon +n\:\beta \delta}}{\partial q^{\alpha }} + \frac{\partial \gamma_{\epsilon +n\:\alpha \beta }}{\partial q^{\delta}} + \frac{\partial \gamma_{\epsilon +n\:\alpha \delta}}{  \partial q^{\beta} } +
2\gamma_{\epsilon +n\:\upsilon \alpha }\gamma_{\upsilon +n\:\beta \delta}+2\gamma_{\epsilon +n\:\upsilon \delta}\gamma_{\upsilon +n\:\alpha \beta }+2\gamma_{\epsilon +n\:\upsilon \beta }\gamma_{\upsilon +n\:\alpha \delta}
\right)
\ee 
and
\be
\label{e7.5}
K_{\alpha \beta \gamma\:\delta +n}= -\frac{1}{3} \Big(K_{\delta +n\:\alpha \beta \gamma}+K_{\delta +n\:\beta \alpha \gamma} \Big),
\ee
\[
K_{\alpha \beta \gamma \delta}= - \frac{1}{3}\sum_{\epsilon ,\upsilon =1}^n p_{\epsilon}  \left(-\frac{\partial K_{\epsilon +n\:\beta \gamma \delta }}{\partial q^{\alpha }} \:
-  \:\frac{\partial K_{\epsilon +n\:\alpha \gamma \delta }}{\partial q^{\beta} }\:+ \:4\gamma_{\epsilon +n\:\alpha \upsilon } K_{\upsilon +n\:\beta \gamma \delta }
\:+4\:\gamma_{\epsilon +n\:\beta \upsilon } K_{\upsilon +n\:\alpha \gamma \delta }+
 \right.
\]
\[
\:-\:\gamma_{\epsilon +n\:\gamma \upsilon } K_{\upsilon +n\:\alpha \beta \delta }
-\gamma_{\epsilon +n\:\gamma \upsilon } K_{\upsilon +n\:\beta \alpha \delta }
+ \gamma_{\epsilon +n\:\delta \upsilon } K_{\upsilon +n\:\alpha \beta \gamma}
+ \gamma_{\epsilon +n\:\delta \upsilon } K_{\upsilon +n\:\beta \alpha \gamma}
+ \gamma_{\upsilon +n\:\alpha \delta } K_{\epsilon +n\:\beta \upsilon \gamma}+
\]
\be
\label{e8}
+ \gamma_{\upsilon +n\:\beta \delta } K_{\epsilon +n\:\alpha \upsilon \gamma}
- \gamma_{\upsilon +n\:\alpha \gamma} K_{\epsilon +n\:\beta \upsilon \delta }
- \gamma_{\upsilon +n\:\beta \gamma} K_{\epsilon +n\:\alpha \upsilon \delta }
-2  \gamma_{\upsilon +n\:\alpha \beta } K_{\epsilon +n\:\upsilon \gamma \delta }
\Big).
\ee
Using (\ref{e7}) we conclude that the  symplectic connection induced by a Riemannian connection is defined by $\frac{1}{2}n^2(n+1)$ functions $\gamma_{\alpha+n\:\beta \delta}$ and their partial derivatives. All of these functions  depend only on spatial coordinates $q^1,  \ldots, q^n.$

Thus from (\ref{e7.5}) and (\ref{e8}) we notice, that  the induced symplectic curvature tensor is fully characterised by its $\frac{1}{3}n^2(n^2-1)$ components: $K_{\alpha+n\:\beta \beta \gamma  },\;  \beta  <\gamma \;,\;
 K_{\alpha+n\:\beta \gamma  \beta },\;  \gamma   <\beta \;,
 K_{\alpha+n\:\beta \gamma  \delta}$ depending only on spatial coordinates, partial derivatives of these coefficients and the symplectic connection coefficients $\gamma_{\alpha +n \: \beta \delta}.$ We remind the reader ( see formula (\ref{e3})) that
\[
K_{\alpha +n\:\gamma  \beta \delta }= K_{\alpha +n\:\beta \gamma  \delta } + K_{\alpha +n\:\delta \beta \gamma  }\;,\;  \beta  <\gamma  <\delta.
\]
It follows from (\ref{e2}), (\ref{e3}) and (\ref{e7.5}) that in any proper Darboux chart
\be
\label{e9}
K_{\alpha \beta \gamma  \:\delta +n}+ K_{\gamma  \alpha \beta \:\delta +n}+K_{\beta \gamma  \alpha \:\delta +n}=0.
\ee
Applying (\ref{e2}), (\ref{e3})  and (\ref{e9}) we get back  (\ref{e7.5}). Thus properties (\ref{e7.5}) and (\ref{e9}) are equivalent.

The immediate consequence of the identity (\ref{e9}) is the relation
$ K_{\alpha \alpha \alpha \:\beta +n}=0.$

In the set  of the symplectic connections fulfilling the requirements: 
\begin{enumerate}
\item
$ \gamma_{\,\alpha +n\:\beta +n\:\delta  }=0,\;\; \gamma_{\,\alpha +n\:\beta +n\:\delta  +n}=0 $,
\item
$\gamma_{\,\alpha +n\:\beta \delta  } $ are some functions of spatial coordinates $q^1, \ldots, q^n ,$
\item
\label{110}
on the base space ${\cal M}$  all of coefficients $\gamma_{\,\alpha \beta \delta  } $ disappear
\end{enumerate}
only the curvature tensor of the induced symplectic connection (\ref{wzor1}) satisfies (\ref{e9}). 
Indeed, for the fixed indices $\alpha ,\beta ,\gamma  ,\delta   $  we obtain from
(\ref{e1}) and (\ref{e9}) that 
\[
0= 
\frac{\partial \gamma_{\alpha \beta \delta +n }}{\partial q^\gamma  }
- \frac{\partial \gamma_{\alpha \beta \gamma}}{\partial p_{\delta}  }+ \omega^{\epsilon \upsilon }\gamma_{\upsilon \alpha
\delta  +n }\gamma_{\epsilon \beta \gamma  }- \omega^{\epsilon \upsilon }\gamma_{\upsilon \alpha \gamma  }\gamma_{\epsilon \beta \delta  +n}+
\frac{\partial \gamma_{\alpha \gamma \delta +n }}{\partial q^{\beta}  }
- \frac{\partial \gamma_{\alpha \beta \gamma }}{\partial p_{\delta} }+ \omega^{\epsilon \upsilon }\gamma_{\upsilon \alpha \delta+n  }\gamma_{\epsilon  \beta \gamma }+
\]
\be
\label{111}
- \omega^{\epsilon \upsilon }\gamma_{\upsilon \alpha \beta   }\gamma_{\epsilon \gamma \delta+n }
+\frac{\partial \gamma_{\beta \gamma \delta+n  }}{\partial q^{\alpha}  }
- \frac{\partial \gamma_{\alpha \beta \gamma  }}{\partial p_{\delta} }+ \omega^{\epsilon \upsilon }\gamma_{\upsilon \gamma \delta +n }\gamma_{\epsilon \alpha \beta  }- \omega^{\epsilon \upsilon }\gamma_{\upsilon \alpha \gamma }\gamma_{\epsilon \beta \delta+n  }.
\ee
Solving the system of differential equations (\ref{111}) numerated by $\delta $ for the fixed $\alpha ,\beta ,\gamma  $ and taking into account the 3rd condition from the list above
 we see that
in Darboux coordinates  every coefficient $\gamma_{ \alpha \beta \gamma  }$ must be of the form (\ref{e7}).

Applying Theorem \ref{ponadto2} to the induced symplectic connection we see that the induced symplectic connection is the only one  compatible symplectic connection which is flat for the flat linear symmetric connection on the base manifold ${\cal M}.$

From (\ref{q0}) and (\ref{e6}) we conclude that potentially nonzero components of the symplectic Ricci tensor are
\[
K_{\alpha \beta}= \frac{1}{3}\Big( R_{\alpha \beta}+ R_{\beta \alpha}\Big),
\]
where $R_{\alpha \beta}$ are elements of the Ricci tensor on the base manifold. In the case when  $R_{\alpha \beta}$ is the Ricci tensor of a Levi-- Civita connection  it is symmetric. Hence a  symplectic manifold ${\cal T^*M}$ is Ricci flat if and only if  the Riemannian  manifold ${\cal M}$ is  Ricci flat.

Assume that the base manifold ${\cal M}$ is a Riemannian manifold   with a metric structure determined locally by the metric tensor $g_{\alpha \beta}. $ This metric structure introduces the unique Levi-- Civita connection on ${\cal M}$ (see formula (\ref{wz1})). A smooth curve $c: {\mathbb R} \rightarrow {\cal M}$ is locally characterized by the system of  equations $q^{\alpha}=q^{\alpha}(t), t \in {\mathbb R}.$ The  {\bf lift} of $c$ from
the Riemannian manifold ${\cal M}$ to the cotangent bundle ${\cal T^* M}$ is the  smooth curve $\tilde{c}: {\mathbb R} \rightarrow {\cal T^*M}$ locally expressed by the set of equations 
\[
\left\{
\begin{array}{ll}
q^{\alpha}=q^{\alpha}(t),&\\
 p_{\alpha}=g_{\alpha \beta }\Big(q^{\delta}(t)\Big) \frac{dq^{\beta}(t)}{dt},& \;t \in {\mathbb R}. 
\end{array}
\right.
\]
Among all symplectic connections compatible with the Levi-- Civita connection on the base manifold
  ${\cal M}$, only in the case of  the induced symplectic connection the lift of any geodesic on ${\cal M}$ with respect to the Levi-- Civita connection on ${\cal M}$ is a geodesic with respect to the induced symplectic connection (\ref{wzor1}) on ${\cal T^* M}$ \cite{pleb1}.

\vspace{1cm}

The second proposed choice  of a compatible symplectic connection can be applied only in certain situations. We shall analyse one of them. 

Let us assume that a $n$--D base manifold ${\cal M}$ can be covered with an atlas in which   all  transition functions are linear. Thus locally  
\[
\forall_{\alpha}\;\; Q^{\alpha}= a^{\alpha}_{\beta}\;q^{\beta}.
\]
By $a^{\alpha}_{\beta}$ we denote elements of the matrix of transformation between `old' coordinates $q^{\beta}$ and `new' ones $Q^{\alpha}.$ The matrix $a$ does not depend on the point and it is nonsingular.

Then in the proper Darboux atlas on the cotangent bundle ${\cal T^*M}$ we  propose the following compatible symplectic connection:
\be
\label{pion1}
\forall_{I,J,K}\;\; \forall_{\alpha, \beta, \delta}\;\;\;
\gamma_{IJK}= \gamma_{IJ\alpha}= \gamma_{\alpha \beta \delta}=0\;\;, \;\; \gamma_{I\alpha \beta}=-\Gamma^{I-n}_{\alpha \beta}(q^1, \ldots, q^n).
\ee
Hence the compatible symplectic connection is completely characterized by its $\frac{1}{2}n^2(n+1)$ coefficients depending only on spatial coordinates. The straightforward consequence of this fact is the observation that among symplectic curvature tensor components  only $K_{I \:\alpha \beta \delta}$  and $K_{\alpha \beta \delta \:I}$ can be
different from $0.$ 
 They are related by    equality (\ref{nowa4}). All of them are functions of spatial coordinates exclusively.


\section{The Fedosov  deformation quantization for a compatible symplectic connection}

In this section we consider some construction of a natural $*$--product on the cotangent bundle ${\cal T^*M}.$ As it is known \cite{sgu} any natural $*$--product on a symplectic manifold $({\cal W}, \omega)$ determines a unique symplectic connection. Moreover, using geometric construction proposed by Fedosov we are able, starting from some fixed symplectic connection on the symplectic manifold, to construct a natural star product on this manifold. Hence we are going to analyse a family of $*$--products constructed according to the Fedosov method and based on compatible symplectic connections.

 All of these $*$--products are equivalent. However, eigenvalues of observables, time evolution of states, mean values of functions etc.  depend in quantum mechanics on the choice of a $*$--product itself instead of its equivalence class. This is the reason why from the physical point of view  the proposed $*$--products seem to be worth analysing.

Considerations presented in this part of our paper have been divided in three subsections. In the first one we analyse the construction and properties of an Abelian connection determined by the compatible symplectic  connections introduced in the second chapter. The next subsection is devoted to  flat sections of the Weyl bundle and 
the $*$-product generated by the compatible  symplectic connections. Some examples end this section.

We assume that the reader is familiar with the Fedosov quantization algorithm. For details see references \cite{6, 7}.

\subsection{The Abelian connection}

Let $({\cal W},\omega,\gamma)$ be a Fedosov manifold covered by an atlas ${\cal A}= \{({\cal U}_z,\phi_z)\}_{z \in J}.$
By $\hbar$ we  denote a deformation   parameter. We assume that it is positive. In physics the deformation parameter $\hbar$ is identified with the Dirac constant. The symbols $y^1, \ldots, y^{2n}$ represent the components of an arbitrary vector ${\bf y}$ belonging to the tangent space $T_{\tt p} {\cal W}$
 at the point ${\tt p} \in {\cal W} $ with respect to   the natural basis $\left( \frac{\partial}{\partial x^i}\right)_{\tt p}$ determined by the chart $({\cal U}_z,\phi_z)$ such that ${\tt p} \in {\cal U}_z.$

We introduce the formal series 
\be
\label{222}
a :=\sum_{l=0}^{\infty}\hbar^k a_{k, j_1 \ldots j_l} y^{j_1} \ldots y^{j_l}, \;\; k \geq 0
\ee
at  point ${\tt p}.$
 By $a_{k, j_1 \ldots j_l}$ we mean the components of a covariant tensor totally symmetric with respect to the indices $\{j_1, \ldots, j_{l}\}$ in  the natural basis $dx^{j_1} \odot \ldots \odot dx^{j_l}. $ For $l=0$ we have $a = \hbar^k a_k$, where  elements $a_k$ are smooth functions on the manifold ${\cal W}.$

The part of the series $a$ standing at $\hbar^k$ and containing $l$ components of the vector ${\bf y}$ will be denoted by
$a[k,l].$ Thus 
\be
\label{2a}
a= \sum_{k=0}^{\infty}\sum_{l=0}^{\infty}\hbar^k a[k,l].
\ee
 The  degree $\deg(a[k,l]) $ of the component $a[k,l]$  equals $2k+l.$

Notice that since $a_{k, j_1 \ldots j_l}$ are totally symmetric in the indices $\{j_1, \ldots, j_l\},$ the element $a$ defined by the formula (\ref{222}) can be understood as the polynomial 
\be
\label{333}
a= \sum_{z=0}^{\infty} \sum_{k=0}^{\left[\frac{z}{2}\right]} \hbar^k \tilde{a}_{k,i_1 \ldots i_{2n}}(y^{1})^{i_1} \ldots 
(y^{2n})^{i_{2n}}, 
\ee
where
$ 0 \leq i_1, \ldots, i_{2n}\leq z-2k \;\;\; , \;\;\;
i_1+ \cdots + i_{2n}= z-2k.$   
\newline
The symbol $\left[\frac{z}{2}\right]$ denotes the integer part of $\frac{z}{2}.$
The relation between the tensor components $a_{k, j_1 \ldots j_l}$ and the polynomial coefficients $\tilde{a}_{k,i_1 \ldots i_{2n}}$
reads
\be
\label{4}
\tilde{a}_{k,i_1 \ldots i_{2n}}= \frac{(z-2k)!}{i_1! \cdots i_{2n}!}\, a_{k\, { \underbrace{ 11 \ldots 1}_{i_1\;\;{\rm indices}} \ldots 
\underbrace{2n \, 2n  \ldots 2n}_{i_{2n} \;\;{\rm indices}} }  }.
\ee

Let $P^*_{\tt p}{\cal W}[[\hbar]]$ denote a set of all elements $a$ of the form (\ref{222}) at the point ${\tt p}. $ \\   
The product  $\circ: P^*_{\tt p}{\cal W}[[\hbar]] \times P^*_{\tt p}{\cal W}[[\hbar]]\rightarrow P^*_{\tt p}{\cal W}[[\hbar]]$ of   two elements $
a, b \in P^*_{\tt p}{\cal W}[[\hbar]]$ is the mapping
\be
\label{5}
a \circ b
:=  \sum_{t=0}^{\infty} \frac{1}{t!}\left(-\frac{i\hbar}{2}\right)^t\omega^{i_1 j_1} \cdots \omega^{i_t j_t} \:\frac{\partial^{ t}  a }{\partial y^{i_1}\ldots\partial y^{i_t} }\:\frac{\partial^{t}  b }{\partial y^{j_1}\ldots\partial y^{j_t} }. 
\ee
As it is known \cite{ja5}, in Darboux coordinates we have
\[
(y^i)^r (y^{i+n})^j \circ (y^i)^s (y^{i+n})^k
= r!\: j!\: s!\: k!\: \sum_{t=0}^{{\rm min}[r,k]+{\rm min.}[j,s]} \left( \frac{i \hbar}{2}\right)^t (y^i)^{r+s-t}(y^{i+n})^{k+j-t}
\times
\]
\be
\label{5.5}
\times \sum_{a={\rm max}[t-r,t-k,0]}^{{\rm min}[j,s,t]}(-1)^a \frac{1}{ a!\: (t-a)!\:(r-t+a)!\:(j-a)!\:(s-a)!\:(k-t+a)!}.
\ee

The pair $(P^*_{\tt p}{\cal W}[[\hbar]],\circ) $ is a noncommutative associative algebra called the  Weyl algebra. 
Taking a set of the Weyl algebras $(P^*_{\tt p}{\cal W}[[\hbar]],\circ) $ at all  points of the manifold ${\cal W}$ we obtain the Weyl bundle 
\[
{\cal P^*W}[[\hbar]] := \bigcup_{{\tt p} \in {\cal W}}   (P^*_{\tt p}{\cal W}[[\hbar]],\circ).
\]

Geometric structure of the Fedosov deformation quantization is based on the 
 $m$-differential form calculus with values in the Weyl bundle. Locally such a form can be written as follows 
\be
\label{9}
a= \sum_{l=0}^{\infty}\hbar^k a_{k, j_1 \ldots j_l, s_1 \ldots s_m}(x^1, \ldots, x^{2n}) y^{j_1} \ldots  y^{j_l}dx^{s_1} \wedge
\cdots \wedge dx^{s_m},
\ee
where $0 \leq m \leq 2n.$ Now
$a_{k, i_1 \ldots i_l, j_1 \ldots j_m}(x^1, \ldots, x^{2n}) $ are components of smooth tensor fields on ${\cal W}$ and 
$ C^{\infty}({\cal TW})\ni {\bf y}\stackrel{\rm locally }{=} y^i \frac{\partial}{\partial x^i} $ is a smooth vector field  on      
${\cal W}.$ We use the same symbol for the vector field 
$  {\bf y} \in  C^{\infty}({\cal TW})$ and the vector ${\bf y} \in  T_{\tt p}{\cal W}.$ 
From now on  we will omit the variables $x^1, \ldots, x^{2n}$ in $a_{k, j_1 \ldots j_l, s_1 \ldots s_m}(x^1, \ldots, x^{2n})$. 

 Differential forms  of the type (\ref{9}) are smooth sections of the direct sum \\
$ {\cal P^*W}[[\hbar]] \otimes \Lambda := \oplus_{m=0}^{2n}({\cal P^*W}[[\hbar]] \otimes \Lambda^m)
$. By $\Lambda^m$ we mean  the space of smooth  $m$--forms on  ${\cal W}.$

The commutator of  forms $a \in C^{\infty}({\cal P^*W}[[\hbar]] \otimes \Lambda^{m_1})$ and  $b \in C^{\infty}({\cal P^*W}[[\hbar]] \otimes \Lambda^{m_2})$ is the form $[a,b] \in C^{\infty}({\cal P^*W}[[\hbar]] \otimes \Lambda^{m_1+m_2})$ defined as
\be
\label{10}
[a,b] := a \circ b - (-1)^{m_1 \cdot m_2}b \circ a.
\ee

\begin{de}
The antiderivation operator 
$
 \delta:C^{\infty}({\cal P^*W}[[\hbar]] \otimes \Lambda^{m}) \rightarrow  C^{\infty}({\cal P^*W}[[\hbar]] \otimes \Lambda^{m+1})
$
 is defined by
\be
\label{e44}
\delta a :=dx^k \wedge \frac{\partial a}{\partial y^k}.
\ee
\end{de}
The operator $\delta$ lowers the degree  of the  elements of  
${\cal P^*W}[[\hbar]] \otimes \Lambda$ by $1$.

Every two forms $a \in C^{\infty}({\cal P^*W}[[\hbar]] \otimes \Lambda^{m_1})$ and $b \in C^{\infty}({\cal P^*W}[[\hbar]] \otimes  \Lambda)$ satisfy
\be
\label{e55}
\delta(a \circ b)= (\delta a)\circ b + (-1)^{m_1} a \circ (\delta b).
\ee

The operator $\delta^{-1}:C^{\infty}({\cal P^*W}[[\hbar]] \otimes \Lambda^m) \rightarrow C^{\infty}({\cal
P^*W}[[\hbar]]
\otimes \Lambda^{m-1})$ is defined by
\be
\label{11}
\delta^{-1} a = \left\{ \begin{array}{ccl}
&\frac{1}{l+m}\: y^k \frac{\partial }{\partial x^k}\rfloor a \qquad  &{\rm for} \;\;\; l+m>0,   \\[0.35cm]
 & 0 \qquad &{\rm for} \;\;\; l+m=0,
\end{array}\right.
\ee
where $l$ is  the degree of $a$ in $y^j$'s and it equals the number of $y^j$'s. 
The operator $\delta^{-1}$ raises the degree of the forms of ${\cal P^*W}[[\hbar]] \otimes\Lambda$ in the Weyl algebra by 
$1$.

The exterior covariant derivative  $\partial_{\gamma}$ of a form $ a \in C^{\infty}({\cal P^*W}[[\hbar]] \otimes \Lambda^m )$ determined by a symplectic connection $\gamma$
 is the linear operator 
\[
\partial_{\gamma} : C^{\infty}({\cal P^*W}[[\hbar]] \otimes \Lambda^m ) \rightarrow C^{\infty}({\cal P^*W}[[\hbar]] \otimes \Lambda^{m+1} ) 
\]
such that
\[
\partial_{\gamma}  a:= dx^k \wedge a_{;k}.
\]
In a Darboux chart 
\be
\label{14}
\partial_{\gamma} a =da + \frac{i}{ \hbar}[\gamma,a].
\ee

The $1$--form   $\gamma$ standing at the commutator
$
\gamma := \frac{1}{2}\gamma_{ijk}y^iy^j dx^k.
$ 
If the connection $\gamma$ is a compatible symplectic connection,
the $1$--form $\gamma$ contains three kinds of elements (the indices $\alpha, \beta, \epsilon,I $ are fixed!):
$\frac{2-\delta^{\beta \epsilon}}{2}\gamma_{I \beta \epsilon}y^{\beta} y^{\epsilon} dp_{I-n} ,\; \beta \leq \epsilon\;,\; 
\gamma_{I\:\beta \epsilon}y^{\beta}y^{I}dq^{\epsilon}$ and $
\frac{2-\delta^{\alpha \beta }}{2}\gamma_{\alpha \beta \epsilon}y^{\alpha}y^{\beta}dq^{\epsilon},\;  \alpha \leq \beta.  $

For every  symplectic connection $1$--form $\gamma$ its antiderivation $\delta \gamma=0.$  Remember that the coefficients $\gamma_{\alpha \beta \epsilon}$  are of the form (\ref{nowa3}).

 The curvature $2$--form $R_{\gamma}$ of  $\gamma$
in a Darboux chart can be expressed by the formula
\be
\label{16}
R_{\gamma}= d \gamma + \frac{i}{2  \hbar}[\gamma, \gamma]=  d \gamma + \frac{i}{  \hbar}\gamma \circ  \gamma. 
\ee
Assume that $\gamma $ is determined by a compatible symplectic connection. Thus we obtain that the form $R_{\gamma}$ consists of three types of terms (all indices are fixed!):
$ \frac{2-\delta^{\alpha \beta}}{2} K_{\alpha \beta \epsilon \: \upsilon+n}y^{\alpha} y^{\beta} dq^{\epsilon} \wedge dp_{\upsilon} \;,\;  \alpha \leq \beta\; ,\;
K_{I \:\alpha \beta \epsilon }y^{\alpha} y^{I}dq^{\beta} \wedge dq^{\epsilon}  \;,\;  \beta < \epsilon , $ and $
 \frac{2-\delta^{\alpha \beta}}{2} K_{\alpha \beta \epsilon \upsilon }y^{\alpha} y^{\beta} dq^{\epsilon} \wedge dq^{\upsilon} ,\;  \alpha \leq \beta \;,\; \epsilon < \upsilon .$
 
The terms $K_{\alpha\beta \epsilon \upsilon}$ are homogeneous functions of momenta $p_{\alpha}.$ 
 Property $\delta R_{\gamma}=0$ follows from (\ref{16}) and (\ref{e55}). It  is equivalent to (\ref{e3}).

Let us introduce a new symbol. The  coefficient $a[\upsilon|i_1, \ldots, i_n|\tau|j,k]$   stands at
\[
a[\upsilon|i_1, \ldots, i_n|\tau|j,k]\;
p_{\upsilon}\: (y^1)^{i_1}\cdots (y^n)^{i_n}y^{\tau+n} dx^j \wedge dx^k, \; 1 \leq j < k \leq 2n. 
\]

The crucial role in  Fedosov's deformation quantization is played by an  Abelian connection $\tilde{\gamma}.$ By definition the Abelian connection $\tilde{\gamma}$ is the connection in the Weyl algebra bundle whose curvature is a central form. Hence for any 
$a \in C^{\infty}({\cal P^*W}[[\hbar]]\otimes \Lambda)$ we obtain $\partial_{\tilde{\gamma}}(\partial_{\tilde{\gamma}}a)=0.$

The Abelian connection proposed by Fedosov is of the form
\be
\label{17}
\tilde{\gamma}= \omega_{ij}y^i dx^j + \gamma + r.
\ee
Its curvature
\[
R_{\tilde{\gamma}}= -\frac{1}{2}\omega_{j_1j_2}dx^{j_1} \wedge dx^{j_2} + R_{\gamma} -\delta r +\partial_{\tilde{\gamma}}+\frac{1}{i \hbar} r \circ r.
\]
The requirement $R_{\tilde{\gamma}}= -\frac{1}{2}\omega_{j_1j_2}dx^{j_1} \wedge dx^{j_2}$ imposes the following condition on the series $r$
\be
\label{17a}
\delta r= R_{\gamma} + \partial_{\gamma} r + \frac{i}{\hbar} r \circ r.
\ee
 Fedosov has shown \cite{6} that  Eq. (\ref{17a}) has a unique solution
\be
\label{18}
r = \delta^{-1} R_{\gamma} + \delta^{-1} \left( \partial_{\gamma} r + \frac{i}{ \hbar}r \circ r \right)
\ee
fulfilling   conditions: $\delta^{-1}r=0$ and $\deg(r) \geq 3.$

Let $r[z]$  denote the  component $r[z]:=\sum_{k=0}^{[\frac{z-1}{4}]}h^{2k}r_{m}[2k,z-4k]dx^m, \;\;z \geq 3$   of $r$   of the degree $z.$ 
Kravchenko \cite{olga} and Vaisman \cite{vais2} found that
\[
r[3]= \delta^{-1}R_{\gamma},
\]
\be
\label{19}
r[z]=\delta^{-1}\left( \partial_{\gamma}r[z-1]+ \frac{i}{ \hbar}\sum_{j=3}^{z-2}r[j] \circ r[z+1-j]\right), \;\; z  \geq 4.
\ee 
This result can be written in a compact form
\be
\label{e20}
r[z]=\delta^{-1} R_{\gamma +r}[z-1],\;\; z \geq 3.
\ee
The expression $R_{\gamma +r}[z-1]$ is the part of the curvature of the connection $\gamma + \sum_{i=3}^{z-1}r[i]$ of the degree $(z-1).$ From the relation (\ref{17a}) we deduce that $\delta R_{\gamma +r}[z]=0$ for $z \geq 2.$ Moreover, the $2$--form  $R_{\gamma +r}$ fulfills the  Bianchi identity
\be
\label{e20.1}
d R_{\gamma +r} + \frac{i}{\hbar}[\gamma +r,R_{\gamma +r}]=0.
\ee

If $\gamma$ is a compatible symplectic connection not containing $\hbar$, there is $\forall_{z \geq 2} \; \frac{\partial R_{\gamma +r}[z]}{\partial \hbar}=0$. Moreover,
we see that $R_{\gamma +r}[z]$ is a sum of three kinds of elements:
\begin{enumerate}
\item
$R[0|i_1, \ldots, i_n|0|\alpha,\beta +n] \:(y^1)^{i_1}\cdots (y^n)^{i_n} dq^{\alpha} \wedge dp_{\beta},$   \vspace{0.2cm} 
$ 0 \leq i_1, \ldots, i_n \leq z, \; i_1 + \cdots + i_n=z.$ 
There are  $n^2 \left( 
\begin{array}{c} z+n-1 \\ z
\end{array}
\right)$ elements of this type.
\item
$R[0|i_1, \ldots, i_n|\tau|\alpha ,\beta]\: (y^1)^{i_1}\cdots (y^n)^{i_n}y^{\tau +n} dq^{\alpha} \wedge dq^{\beta},$   \vspace{0.2cm} 
$ 0 \leq i_1, \ldots, i_n \leq z-1, \; i_1 + \cdots + i_n=z-1,  \; \alpha < \beta.$
We get $n \left( 
\begin{array}{c} (z-1)+n-1 \\ z-1
\end{array}
\right) \left( 
\begin{array}{c} n \\ 2
\end{array}
\right) $ terms of this form. 
\item
$R[\upsilon|i_1, \ldots, i_n|0|\alpha ,\beta] \:p_{\upsilon} \:(y^1)^{i_1}\cdots (y^n)^{i_n} dq^{\alpha} \wedge dq^{\beta},$  \vspace{0.2cm} 
$ 0 \leq i_1, \ldots, i_n \leq z, \; i_1 + \cdots + i_n=z,  \; \alpha < \beta.$ There are  
$n \left( 
\begin{array}{c} z+n-1 \\ z
\end{array}
\right) \left( 
\begin{array}{c} n \\ 2
\end{array}
\right) $  elements of this kind. Every function $R[\upsilon |i_1, \ldots, i_n|\tau |j,k]$ depends only on spatial coordinates $q^1, \ldots, q^n.$
\end{enumerate}

There are some constraints imposed on these functions.
 All  elements from the first class are chosen to be independent. Every coefficient standing at a term from the second group is determined by two coefficients belonging to  the first class (see formula (\ref{31})). Among the third set  we can choose a special group of 
$n(z+1)\left(
\begin{array}{c}
n+z\\
z+2
\end{array}\right)$  
elements. The selection method will be  presented below. Any other coefficient from the third group is a linear function of these selected ones.
\vspace{0.2cm}

Let us consider consequences of the restriction $\delta R_{\gamma+r}[z]=0, \; z \geq 2.$
We start from \\ $n \left( \begin{array}{c}
(z-1)+n-1 \\
z-1
\end{array}
\right) \left( 
\begin{array}{c} n \\ 2
\end{array}
\right) $ nontrivial equations containing terms of the first and of the second kind.
\[
\Big(
(i_{\alpha}+1)R[0|i_1, \ldots,i_{\alpha}+1, \ldots,i_{\beta}, \ldots, i_n|0|\beta,\tau+n] - (i_{\beta}+1) R[0|i_1, \ldots,i_{\alpha}, \ldots,i_{\beta}+1, \ldots, i_n|0|\alpha,\tau+n]+ 
\]
\be
\label{30}
+ R[0|i_1, \ldots,i_{\alpha}, \ldots,i_{\beta}, \ldots, i_n|\tau|\alpha,\beta] \Big)
\:(y^1)^{i_1}\cdots(y^{\alpha})^{i_{\alpha}}\cdots (y^{\beta})^{i_{\beta}}\cdots  (y^n)^{i_n} dq^{\alpha} \wedge dq^{\beta} \wedge dp_{\tau}=0,  
\ee
where $i_1 + \cdots + i_n=z-1, \:  \alpha < \beta.$
 The number of conditions (\ref{30}) equals the number of coefficients of the second type. Moreover, in each  equation  only one   coefficient of the type $R[0|i_1, \ldots,i_{\alpha}, \ldots,i_{\beta}, \ldots, i_n|\tau|\alpha,\beta]$ 
appears
and each term  
$R[0|i_1, \ldots,i_{\alpha}, \ldots,i_{\beta}, \ldots, i_n|\tau|\alpha,\beta]$ is present in exactly one of these equations.
Hence we conclude that every coefficient from the second set can be uniquely expressed by elements from the first collection. Indeed, from (\ref{30})
\[
R[0|i_1, \ldots,i_{\alpha}, \ldots,i_{\beta}, \ldots, i_n|\tau|\alpha,\beta]= 
(i_{\beta}+1) R[0|i_1, \ldots,i_{\alpha}, \ldots,i_{\beta}+1, \ldots, i_n|0|\alpha,\tau+n]+
\]
\be
- (i_{\alpha}+1)R[0|i_1, \ldots,i_{\alpha}+1, \ldots,i_{\beta}, \ldots, i_n|0|\beta,\tau+n],\;\;
\label{31}
  0 \leq i_1, \ldots, i_n \leq z-1, \; i_1 + \cdots + i_n=z-1, \;  \alpha < \beta.
\ee
The conditions
\[
\Big( (i_{\alpha} +1) R[0|i_1,\ldots, i_{\alpha} +1,\ldots,i_{\beta}, \ldots,i_{\kappa}, \ldots, i_n|\tau|\beta,\kappa] - (i_{\beta}+1) R[0|i_1,\ldots, i_{\alpha}, \ldots,i_{\beta}+1, \ldots,i_{\kappa}, \ldots, i_n|\tau|\alpha ,\kappa] +  
\]
\[
+(i_{\kappa}+1)  R[0|i_1,\ldots, i_{\alpha}, \ldots,i_{\beta}, \ldots,i_{\kappa}+1, \ldots, i_n|\tau|\alpha ,\beta]\Big)\times
\]
\[
 \times
  (y^1)^{i_1}\cdots (y^{\alpha} )^{i_{\alpha} }\cdots(y^{\beta})^{i_{\beta}}\cdots (y^{\kappa})^{i_{\kappa}}\cdots  (y^n)^{i_n} y^{\tau+n} dq^{\alpha}  \wedge dq^{\beta} \wedge dq^{\kappa}   =0,
 \]
\be
\label{32}
0 \leq i_1, \ldots, i_n \leq z-2, \; i_1+ \cdots + i_n=z-2, \;  \alpha  <\beta < \kappa 
\ee
are nontrivial for $3 \leq n.$ They 
are imposed only on terms of the second kind. However, applying   relations  (\ref{31})
we turn them into identities.

Moreover, $n^2$ coefficients  of the type $R[0|0,\ldots,0, i_{\alpha}=z, 0,\ldots,0|0|\alpha,\tau+n]$ disappear because
\be
\label{32.5}
\delta \Big(R[0|0,\ldots,0, i_{\alpha}=z, 0,\ldots,0|0|\alpha,\tau+n](y^{\alpha})^z dq^{\alpha} \wedge dp_{\tau} \Big)=0.
\ee

Finally, let us investigate  consequences of the condition $\delta R_{\gamma+r}[z]=0, \; z \geq 2$ for the elements of the third type. 
As  can be easily checked, each of these  elements appears  in  at most ${\rm min}[z,n-2]$ equations.

 Assume that the  element $R[\upsilon|i_1, \ldots, i_n|0|\alpha,\beta] \:p_{\upsilon} \:(y^1)^{i_1}\cdots (y^n)^{i_n} dq^{\alpha} \wedge dq^{\beta},$ is present exactly in $f$ equations, $ 0 \leq f \leq {\rm min}[z,n-2]$  following from the general condition $\delta R_{\gamma+r}[z]=0.$ It means
that among all possible indices $i_{\kappa}$ there are exactly $f$ numbers $i_{\eta}$ such that $\eta \neq \alpha, \eta \neq \beta$ and $ i_{\eta} \neq 0.$
The total number of terms present in  exactly $f$ equations each equals
$
n \left(
 \begin{array}{c} n \\ 2
\end{array}
\right) 
\left(
\begin{array}{c} n-2 \\ f
\end{array}
\right)
\left(
\begin{array}{c} z+1 \\ z-f
\end{array}
\right).  
$
 
Each of the constraints following from the condition  $\delta R_{\gamma+r}[z]=0$ contains three different coefficients $R[\upsilon|i_1, \ldots, i_n|0|\alpha, \beta]$ and each of these coefficients appears in the same number of equations following from the constraint $\delta R_{\gamma+r}[z]=0$. Moreover, any arbitrary  pair of  coefficients appears in at most one equation.
 
Hence the set of 
$
n \left(
 \begin{array}{c} n-1+(z-1) \\ z-1
\end{array}
\right) 
\left(
\begin{array}{c} n \\ 3
\end{array}
\right)
$
equations following from the requirement $\delta R_{\gamma+r}[z]=0$ can be divided in ${\rm min}[z,n-2]$ separate classes
containing only coefficients $R[\upsilon|i_1, \ldots, i_n|0|\alpha,\beta]$ appearing in the analysed formulas for exactly $f$ times each, where $  1 \leq f \leq {\rm min}[z,n-2].$ Each class consists of 
$
n \left(
 \begin{array}{c} n \\ f+2
\end{array}
\right) 
\left(
\begin{array}{c} z+1 \\ z-f
\end{array}
\right)
$
independent blocks. Every block is a system of 
$
\frac{f}{3}  \left(
 \begin{array}{c} f+2 \\ 2
\end{array}
\right) 
$
linear equations containing $\left(
 \begin{array}{c} f+2 \\ 2
\end{array}
\right) $ coefficients. 
Among these equations only $\left(
 \begin{array}{c} f+1 \\ 2
\end{array}
\right) $  are linearly independent. Therefore from  
$\left(
 \begin{array}{c} f+2 \\ 2
\end{array}
\right) $ coefficients only
$(f+1)$ terms are independent. The choice of these $(f+1)$ elements is not arbitrary. We propose it below.

Each expression $R[\upsilon|i_1, \ldots, i_n|0|\alpha,\beta]\:p_{\upsilon} \:(y^1)^{i_1}\cdots (y^n)^{i_n} dq^{\alpha} \wedge dq^{\beta}$ belongs to exactly one block so each block may be characterized by the quantity $p_{\upsilon} \:(y^1)^{i_1}\cdots (y^n)^{i_n} dq^{\alpha} \wedge dq^{\beta}$. However, the block  $p_{\upsilon} \:(y^1)^{i_1}\cdots (y^n)^{i_n} dq^{\alpha} \wedge dq^{\beta}$ can be equivalently  determined  
 by the expression  
$p_{\upsilon} \:(y^1)^{i_1}\cdots (y^{\alpha})^{i_{\alpha}+1} \cdots (y^{\beta})^{i_{\beta}+1} \cdots(y^n)^{i_n}$. 
In the next paragraph
we use this latter characterization of blocks.

Let us  consider the block
 $p_{\upsilon} (y^{s_1})^{i_{s_1}}\cdots (y^{s_{f+2}})^{i_{s_{f+2}}}$,
 where  the indices satisfy the conditions  $ \forall_{ 1 \leq l \leq f+2}\; 1 \leq i_{s_l}$ and $ 1 \leq s_1<s_2< \ldots <s_{f+2} \leq n $.
As the independent $f+1$ coefficients $R[\upsilon|i_1, \ldots, i_n|0|\alpha,\beta]$ we choose the elements
standing at the exterior products
$dq^{s_1} \wedge dq^{s_{f+2}}, \ldots, dq^{s_{f+1}} \wedge dq^{s_{f+2}}.$  \label{indep}
After simple but tedious calculations we arrive at the following relation:
\[
R[\upsilon|\ldots, i_{s_{j}}-1, \ldots,i_{s_{k}}-1,\ldots,i_{s_{f+2}},\ldots|0|s_{j}, s_{k}]= 
\frac{i_{s_{k}}}{i_{s_{f+2}}}
R[\upsilon|\ldots, i_{s_{j}}-1, \ldots,i_{s_{k}},\ldots,i_{s_{f+2}}-1,\ldots|0|s_{j}, s_{f+2}]+ 
\]
\be
\label{okr1}
-
\frac{i_{s_{j}}}{i_{s_{f+2}}}
R[\upsilon|\ldots, i_{s_{j}},\ldots,i_{s_{k}}-1, \ldots,i_{s_{f+2}}-1,\ldots|0|s_{k}, s_{f+2}]. 
\ee 
The formula (\ref{okr1}) can be applied in the cases when $j<k<f+2.$

What is amazing, the elements $R[\upsilon|i_1, \ldots, i_n|0|\alpha ,\beta], \;\; \sum_{j=1}^n i_j=z$ are determined by  $R[0|i_1, \ldots, i_n|0|\alpha,\beta+n]$ i.e. the elements of the first type and the Abelian connection components of the degree less than $z$. The explicit form of this relation is contained in  the Appendix.

\vspace{0.2cm}
From (\ref{e20}) we compute the element $r[z+1],\; 2 \leq z$. We use the notation analogous to that  applied in the previous considerations  and by $r[\upsilon|i_1, \ldots, i_n|\tau|j]$ we mean the coefficient standing at
$
p_{\upsilon}\: (y^1)^{i_1}\cdots (y^n)^{i_n}y^{\tau+n} dx^j, \;\; 1 \leq j  \leq 2n. 
$
\label{ola}

There are three kinds of components of $r[z+1]$:
\begin{enumerate}
\item
$r[0|i_1, \ldots, i_n|0|\alpha +n] \:(y^1)^{i_1}\cdots (y^n)^{i_n} dp_{\alpha},$ 
\vspace{0.2cm}
\\
$ 0 \leq i_1, \ldots, i_n \leq z+1, \; i_1 + \cdots + i_n=z+1.$ 
We have   $n \left( 
\begin{array}{c} z+n \\ z+1
\end{array}
\right)$ elements of this type. They appear as images of the elements $R_{\gamma+r}[z]$ of the first kind in the mapping $\delta^{-1}.$ 
\item
$r[0|i_1, \ldots, i_n|\tau|\alpha] \:(y^1)^{i_1}\cdots (y^n)^{i_n}y^{\tau+n} dq^{\alpha},$ 
\vspace{0.2cm}
\\
$ 0 \leq i_1, \ldots, i_n \leq z, \; i_1 + \cdots + i_n=z.$ 
The number of expressions of this form is   $n^2 \left( 
\begin{array}{c} z+n-1 \\ z
\end{array}
\right)$. They come from applying the $\delta^{-1}$ operator to the terms $R_{\gamma+r}[z]$ of the first and the second type. And finally
\item
$r[\upsilon|i_1, \ldots, i_n|0|\alpha] \:p_{\upsilon} (y^1)^{i_1}\cdots (y^n)^{i_n} dq^{\alpha},$ 
\vspace{0.2cm}
\\
$ 0 \leq i_1, \ldots, i_n \leq z+1, \; i_1 + \cdots + i_n=z+1.$ There are
$n^2 \left( 
\begin{array}{c} z+n \\ z+1
\end{array}
\right)$ elements  generated by components of  $R_{\gamma+r}[z]$ of the third type.
\end{enumerate}
The same classification can be applied for the $1$-form of the symplectic connection $\gamma$.
The total symmetry of components $\gamma_{ijk}$ in the indices $\{i,j,k\}$  implies 
\be
\label{1203101}
\forall_{\alpha, \beta, \tau} \;\;\;\gamma[0|0, \ldots, i_{\alpha}=1,0, \ldots|\tau|\beta]= 
\gamma[0|0, \ldots, i_{\beta}=1,0, \ldots|\tau|\alpha],
\ee
\be 
\label{1203102}
\forall_{\alpha,  \tau} \;\; \gamma[0|i_1, \ldots, i_{\alpha}, \ldots, i_n|\tau|\alpha]= (i_{\alpha}+1)
\gamma[0|i_1, \ldots, i_{\alpha}+1, \ldots,i_n|0|\tau+n],
\ee
\be
\label{1203103}
\forall_{\alpha <  \beta} \;\; \forall_{\upsilon }  \;\;0 < i_{\beta} \;\;\gamma[\upsilon|i_1, \ldots, i_{\alpha}, \ldots,i_{\beta}, \ldots,i_n|0|\alpha]= \frac{i_{\alpha}+1}{i_{\beta}} \gamma[\upsilon|i_1, \ldots, i_{\alpha}+1, \ldots,i_{\beta}-1, \ldots,i_n|0|\beta].
\ee

Let us consider some relations between three classes of components of $r[z+1], \; 2 \leq z.$ 

After simple  calculations we conclude that
\be
\label{e33}
 r[0|\ldots, i_{s_1}, \ldots, i_{s_u}, \ldots|0|\alpha+n]=\frac{1}{z+2} \sum_{l=1}^u R[0|
\ldots,i_{s_1}, \ldots,i_{s_l}-1, \ldots, i_{s_u}, \ldots|0|s_l,\alpha+n ].
\ee
We assume that $1 \leq  u \leq n,\;\;\forall_l \;0 < i_{s_l}, \;\; i_{s_1}+ \cdots+ i_{s_u}=z+1.  $

In the special case 
$
r[0|0,\ldots, i_{\beta}=z+1,0, \ldots|0|\alpha+n]= \\ =\frac{1}{z+2}  R[0|0, \ldots,(i_{\beta}-1)=z,0,\ldots|0|\beta,\alpha+n ].
$
\vspace{0.2cm}
\newline
But we know that $\forall_{\beta} \;R[0|0, \ldots,(i_{\beta}-1)=z,0,\ldots|0|\beta,\alpha+n ]=0 $ so 
\be
\label{e33.5}
r[0|0,\ldots, i_{\beta}=z+1,0, \ldots|0|\alpha+n]=0.
\ee

The elements of the second kind are determined by the formula
\[
 r[0|\ldots, i_{s_1}, \ldots, i_{s_u}, \ldots|\tau|\alpha]
\stackrel{\rm (\ref{31})}{=}
 \frac{i_{\alpha}-z-1}{z+2}R[0|\ldots, i_{s_1}, \ldots, i_{s_u}, \ldots|0|\alpha,\tau+n]+
\]
\be
\label{e35}
+\frac{i_{\alpha}+1}{z+2}
\sum_{{\rm all }\;l\;{\rm that}\:s_l \neq \alpha}  R[0|\ldots, i_{s_1},\ldots, i_{s_l}-1, \ldots,i_{\alpha}+1, \ldots, i_{s_u}, \ldots|0|s_{l},\tau+n], 
\ee
\[
i_{s_1}+ \cdots + i_{s_u}=z. 
\]
Remember that for all $l$ there is $s_l \neq \alpha $ and $ 0 <s_l$ but it may be $i_{\alpha}=0.$

The straightforward consequence of the relation (\ref{e35}) is the statement that for all $ \alpha, \tau $
\be
\label{e35.5}
r[0,\ldots,0,i_{\tau}=z,0,\ldots,0|\alpha|\tau]=0.
\ee

Applying (\ref{e33}) to the result (\ref{e35}) we obtain
\[
r[0|\ldots, i_{s_1}, \ldots, i_{s_u}, \ldots|\tau|\alpha]=
\]
\be
\label{9032010}
= (i_{\alpha}+1)r[0|\ldots, i_{s_1}, \ldots,i_{\alpha}+1, \ldots, i_{s_u}, \ldots|0|\tau+n]-
R[0|\ldots, i_{s_1},\ldots, i_{s_u}, \ldots|0|\alpha, \tau+n].
\ee
Finally we present formulas determining components of the correction $r$ belonging to the 3rd category. 
\[
r[\upsilon|\ldots, i_{s_1}, \ldots, i_{s_u}, \ldots|0|\alpha]= \frac{1}{z+2}
\sum_{{\rm all}\;l \;{\rm that\:}s_l<\alpha}R[\upsilon|\ldots, i_{s_1},\ldots, i_{s_l}-1, \ldots,i_{\alpha}, \ldots, i_{s_u}, \ldots|0|s_l,\alpha]+
\]
\be
\label{e36}
-
\frac{1}{z+2} \sum_{{\rm all}\;l \;{\rm that\:}s_l>{\alpha}}R[\upsilon|\ldots, i_{s_1},\ldots,i_{\alpha}, \ldots, i_{s_l}-1, \ldots, i_{s_u}, \ldots|0|\alpha,s_l]. 
\ee
The straightforward consequence of (\ref{e36}) is the equality
\be
\label{e36.4}
r[\tau|0,\ldots,0, i_{\alpha}=z+1,0, \ldots, 0|0|\alpha]=0 
\ee
which also results from the condition $\delta^{-1}r=0.$
Relations (\ref{e33.5}),  (\ref{e35.5}) and  (\ref{e36.4}) yield (\ref{32.5}).

Not all  elements $R[\upsilon|\ldots, i_{s_1}, \ldots, i_{s_u}, \ldots|0|\alpha,\beta]$ are independent. Assume that $s_u=\alpha$ in (\ref{e36}). Then
\be
\label{e36.5}
r[\upsilon|\ldots, i_{s_1}, \ldots, i_{s_u}, \ldots|0|\alpha]= \frac{1}{z+2} \sum_{l=1}^{u-1}
R[\upsilon|\ldots, i_{s_1},\ldots, i_{s_l}-1, \ldots, i_{s_u}, \ldots|0|s_l,\alpha].
\ee
The relation determining $r[\upsilon|\ldots, i_{s_1}, \ldots, i_{s_u}, \ldots|0|\alpha]$ for $s_u<\alpha$ is a slight modification of
(\ref{e36.5})
\be
\label{e36.6}
r[\upsilon|\ldots, i_{s_1}, \ldots, i_{s_u}, \ldots|0|\alpha]= \frac{1}{z+2} \sum_{l=1}^{u}
R[\upsilon|\ldots, i_{s_1},\ldots, i_{s_l}-1, \ldots, i_{s_u}, \ldots|0|s_l,\alpha].
\ee
For $s_u>\alpha$ from (\ref{okr1}) and (\ref{e36.5}) we obtain
\[
r[\upsilon|\ldots, i_{s_1}, \ldots, i_{s_u}, \ldots|0|\alpha]= \frac{i_{\alpha}+1}{i_{s_u}} r[\upsilon|\ldots, i_{s_1}, \ldots,i_{\alpha}+1,\ldots, i_{s_u}-1, \ldots|0|s_u]+
\]
\be
\label{e37}
- \frac{1}{i_{s_u}} R[\upsilon|\ldots, i_{s_1}, \ldots,i_{\alpha}, \ldots, i_{s_u}-1, \ldots|0|\alpha,s_u].
\ee
Remember, that although the formulas (\ref{e36.5}),(\ref{e36.6}) and (\ref{e37}) contain the curvature $2$--form components of the third kind, due to the  Bianchi identity they are in fact determined exclusively by the elements $R[0|i_1, \ldots, i_n|0|\alpha, \beta +n].$

Then we are ready to construct the iterative formula determining $R_{\gamma+r}[z]$ by all $R_{\gamma+r}[v], \; 2 \leq v \leq z-1.$ We see that it is sufficient to find the relation describing components $R_{\gamma+r}[z]$ of the first kind.

 Starting from the definition of the curvature $R_{\gamma + r}$ and   applying the formula (\ref{5.5}) we obtain that
\[
R[0|i_1, \ldots, i_n|0|\alpha,\beta+n]= \frac{\partial}{\partial q^{\alpha}}r[0|i_1, \ldots, i_n|0|\beta+n]-
r[\beta|i_1, \ldots, i_n|0|\alpha]+ \sum_{m=1}^n \Bigg(
\] 
\[
 \sum_{\stackrel{ g_1+\ldots +g_n=1}{\forall_c 0 \leq g_c \leq {\rm min}[i_c,1] }}(g_m+1)\cdot
\gamma[0|g_1, \ldots, g_m+1, \ldots,g_n|0|\beta+n]  \cdot r[0|i_1-g_1,\ldots, i_m-g_m, \ldots, i_n-g_n|m|\alpha]+
\]
\[
+  \sum_{\stackrel{g_1+\ldots +g_n = z-1}{\forall_c {\rm max}[0,i_c-1] \leq g_c \leq {\rm min}[i_c,z-1] }}(g_m+1)
r[0|g_1, \ldots, g_m+1, \ldots,g_n|0|\beta+n]  
 \gamma[0|i_1-g_1,\ldots, i_m-g_m, \ldots, i_n-g_n|m|\alpha]+
\]
\be
\label{e38}
+  \sum_{\stackrel{2 \leq g_1+\ldots +g_n \leq z-2}{\forall_c 0 \leq g_c \leq {\rm min}[i_c,z-2] }}(g_m+1)\cdot
r[0|g_1, \ldots, g_m+1, \ldots,g_n|0|\beta+n]  \cdot r[0|i_1-g_1,\ldots, i_m-g_m, \ldots, i_n-g_n|m|\alpha] \Bigg).
\ee
 The sum $i_1+ \ldots + i_n=z.$
This iterative formula
plus relations presented before completely define the Fedosov $*$-product with the symplectic connection compatible with some linear connection according to the scheme proposed in the second Section. We see that to construct the Abelian connection it is sufficient to know the symplectic connection coefficients $\gamma_{I \alpha \beta }$ and the symplectic curvature tensor components $K_{\alpha \beta \delta I}.$ The coefficients $\gamma_{ \alpha \beta \delta}$ influence the Abelian connection $\gamma + r$ only indirectly through the elements $K_{\alpha \beta \delta I}.$ 

In fact 
the  Fedosov scheme of calculating the Abelian connection for a compatible symplectic connection reduces to the loop:
\begin{enumerate}
\item
the Abelian connection $1$--form $\gamma + \sum_{l=3}^{z} r[l]$ and 
the curvature $2$--form  $\sum_{l=2}^{z-1}R_{\gamma +r}[l] $ elements of the 1st and 3rd type 
 are known;
 \item
 from (\ref{e38}) one gets all components of the curvature $2$--form of the 1st type \\ $R[0|i_1, \ldots, i_n|0|\alpha,\beta +n], \;\; i_1 + \cdots + i_n=z$ ;
 \item
then from the formula presented in the Appendix  one has 
$R[\upsilon|\ldots, i_{s_1}, \ldots, i_{s_u}, \ldots|0|\alpha, \beta]$ for $  s_u \leq \beta
, \;\; i_{s_1} + \cdots + i_{s_u}=z$;
\item
using the formula (\ref{e33}) one finds all of the coefficients $r[0|i_1, \ldots, i_n|0|\alpha +n], \;\; i_1 + \cdots + i_n=z+1;$
\item
 from (\ref{9032010}) one obtains the  elements of the form
$r[0| i_1, \ldots, i_n|\tau|\alpha], \;\; i_1 + \cdots + i_n=z.$
\item
in the next step one calculates $r[\upsilon| i_1, \ldots, i_n|0|\alpha], \;\; i_1 + \cdots + i_n=z+1$ applying  relations
(\ref{e36.5}), (\ref{e36.6}) and (\ref{e37}); 
\item
one  comes back to the 1st step of the loop with $z \rightarrow z+1$.
\end{enumerate}

\subsection{The $*$- product of functions}

Let $({\cal W},\omega, \gamma)$ be a Fedosov manifold and ${\cal P^*W}[[\hbar]]$ the Weyl algebra bundle over the manifold 
$({\cal W},\omega, \gamma).$ The Weyl bundle is equipped with the Abelian connection $\tilde{\gamma} $ of the form (\ref{17}).

The subalgebra  ${\cal P^*W}[[\hbar]]_{\tilde{\gamma}}  \subset C^{\infty}( {\cal P^*W}[[\hbar]]  )$ consists of flat sections of the Weyl bundle $C^{\infty} ({\cal P^*W}[[\hbar]]) $ i.e. the sections satisfying the condition
$
\forall a \in {\cal P^*W}[[\hbar]]_{\tilde{\gamma}}\;
\partial_{\tilde{\gamma}}a=0.
$
Fedosov  showed \cite{6, 7} that for any $a_0 \in C^{\infty}(\cal{W}) $ there exists a unique  $a \in {\cal P^*W}[[\hbar]]_{\tilde{\gamma}} $ such that  
\[
 \sigma(a):=a|_{{\bf y}=0}=a_0.
\] 
The element $a= \sigma^{-1}(a_0)$ can be found by the iteration
\be
\label{20}
a= a_0 + \delta^{-1} \left( \partial_{\gamma}a + \frac{i}{ h}[r,a]\right).
\ee
This relation means that
\[
a[0]=a_0,
\]
\be
\label{21}
a[z]=\delta^{-1} \Big( \partial_{\gamma}a[z-1]+\frac{i}{ h} \sum_{l=1}^{z-2} \Big[r[z+1-l],a[l]\Big]\Big), \;\;\;   z \geq 1.
\ee

Hereafter we restrict our considerations to the situation where the symplectic manifold $({\cal W},\omega)$ is a cotangent bundle
${\cal T^*M}$ and $\gamma$ is a compatible symplectic connection on it.
We focus on some properties of the series $\sigma^{-1}(a_0)$ in this case. 

 Let us start from the following observation.
\begin{co}
\label{co23}
An element $\hbar^k g\:(y^1)^{i_1}\cdots (y^{2n})^{i_{2n}} \in C^{\infty}({\cal P^*W}[[\hbar]]),\;g \in C^{\infty}({\cal T^*M}), \; i_1+ \cdots + i_{2n} \geq 1$ is given. The expression
\[
\delta^{-1}\left( \frac{i}{\hbar}\Big[\gamma + r, \hbar^k g \:(y^1)^{i_1}\cdots (y^{2n})^{i_{2n}}\Big]\right) =   
\sum_{d=0}^{\infty}\: \sum_{ j_1+\cdots+j_{2n}=1}^{\infty} 
\hbar^{k+2d} b_{k+2d\: j_1 \ldots j_{2n}} \: 
(y^1)^{j_1}\cdots (y^{2n})^{j_{2n}},
\]
where $\gamma$ is the $1$--form representing a compatible symplectic connection, $r$ is the Abelian connection series generated by $\gamma$, and $b_{k+2d\: j_1 \ldots j_{2n}}$ are  some smooth functions on  ${\cal T^*M}.$ Moreover
\[
\sum_{l=1}^{n}j_{l+n}= \sum_{l=1}^{n}i_{l+n}-2d
\]
for elements obtained from commutators  with $\gamma$ or $r$ of the first and of the second kind (see the classification in the previous subsection) 
and
\[
\sum_{l=1}^{n}j_{l+n}= \sum_{l=1}^{n}i_{l+n}-2d-1
\]
if components of $\gamma$ or $r$ were of the third kind.
\end{co}
This corollary is the straightforward consequence of two definitions: of  the commutator and of the operator $\delta^{-1}.$

Thus we see that the commutators appearing in the recurrence (\ref{20}) do not increase the number of $y$'s with momenta indices. Moreover $\forall_{K}\;\: j_K \leq i_K+1$ if the commutators are calculated with elements of $\gamma + r$ of the $1$st and $2$nd kind and  $\forall_{K}\;\: j_K \leq i_K$ if the commutators with $\gamma + r$ of the $3$rd type are considered. We recall that the capital letters correspond to momenta coordinates.

Therefore we observe that the total number of momenta-like elements
$(y^{K})^{i_K}$ may increase only in  the operation $\delta^{-1}\Big( \hbar^k d g\:(y^1)^{i_1}\cdots (y^{2n})^{i_{2n}}\Big). $ 

In contrary, let   an element $g\:(y^1)^{i_1}\cdots (y^{2n})^{i_{2n}},\; g \in C^{\infty}({\cal T^*M}),  \; \sum_{s=1}^n i_s=l$ be given. Then  every term generated from this element by the recurrence (\ref{20}) contains {\bf at least} $l$ position-like components $y^{\alpha}.$

\begin{co}
\label{co2}
Let $a_0= (p_1)^{i_1}\cdots ({p_n})^{i_n}f(q^1, \ldots, q^n)$ be a smooth function defined on the cotangent bundle ${\cal T^*M}.$ Then $\sigma^{-1}(a_0)$ consists only of elements of the form
\[
\hbar^{2d}  g_{2d\:l_1 \ldots l_n\; j_1 \ldots j_{2n}} (q^1, \ldots, q^n) (p_1)^{l_1} \cdots (p_n)^{l_n}      
(y^1)^{j_1}\cdots (y^{2n})^{j_{2n}}
\]
such that 
$
2d+ \sum_{s=1}^n l_{s}+   \sum_{s=1}^n  j_{n+s} = \sum_{s=1}^n i_s. 
$
\end{co}
The proof of this corollary can be done with the use of the mathematical induction and  Corollary \ref{co23}.

Hence we conclude that if $a_0$ is a function   of the spatial coordinates only then the series $\sigma^{-1}(a_0)$ contains neither  positive powers of the deformation parameter $\hbar$ nor $y^K.$ Thus $\sigma^{-1}(a_0)[z]$ is a polynomial in $y^{\alpha}$ of the degree $z.$ Moreover,
there are not any terms $(p_l)^{l_v}, \;\ l_v >0$ in this series. Degrees of partial derivatives of $a_0$ in $\sigma^{-1}(a_0)[z]$ are from the ordered  set $\{1, \ldots, z\}.$

In the process of generating $\sigma^{-1}(a_0)$ in this case only two kinds of elements can be different from $0$: the exterior derivatives and the commutators with components $\gamma + r$ of the $2$nd category. 

For any  function $a_0 \in C^{\infty}({\cal T^*M})$ the element $a[z], z \geq 1$ consists  of the terms of the form
\[
\hbar^{2d} (p_1)^{l_1}\ldots (p_n)^{l_n} f(q^1, \ldots, q^n)\frac{\partial^{i_1 +\cdots +i_{2n}}a_0}{\partial^{i_1}q^1\ldots \partial^{i_{2n}}p_{n} }(y^1)^{j_1}\cdots (y^{2n})^{j_{2n}}.
\]
This is a short list of properties of the above expressions.
\begin{enumerate}
\item
$4d+j_1+ \cdots +j_{2n}=z\;\;,\;\;1 \leq j_1 + \cdots +j_{2n} \leq z \;\;,\;\;
1 \leq i_1 +\cdots +i_{2n} \leq z,$
\item
\be
\label{kiedy}
\sum_{s=1}^{n}i_{s+n}=2d+ \sum_{s=1}^{n}j_{s+n}+ \sum_{s=1}^{n}l_s.
\ee
 Corollary \ref{co2} is compatible with this statement. Moreover, if the function $a_0$ depends only on momenta $p_1, \ldots, p_n$ then for elements $\sigma^{-1}(a_0)$ containing exclusively momenta--like components $y^K$ there is
\[
\sum_{s=1}^{n}i_{s+n}=2d+ \sum_{s=1}^{n}j_{s+n}.
\] 
\item The maximal value of the sum $\sum_{s=1}^{n}l_s=\left[ \frac{z}{2}\right].$ If $\sum_{s=1}^{n}l_s= \frac{z}{2},$ where $z $ is an even number, then 
$2d+ \sum_{s=1}^{n}j_{s+n}=0$ and $\sum_{s=1}^{n}i_{s+n}= \frac{z}{2}.$
\item
Every function $f(q^1, \ldots, q^n)$ is a polynomial in symplectic connection coefficients $\gamma_{I \alpha \beta},$ partial derivatives of components $\gamma_{\alpha \beta \delta}$ with respect to momenta and partial derivatives of quantities from these two groups with respect to spatial coordinates. 
\end{enumerate}
All of these observations follow from a simple but boring analysis of the Fedosov recurrence.
\vspace{0.5cm}

Using the one-to-one correspondence between the collection of the flat sections ${\cal P^*W}[[\hbar]]_{\tilde{\gamma}}$ and the set  $C^{\infty}(\cal{W})$ we introduce the associative star product `$*$' of functions $a_0,b_0 \in C^{\infty}(\cal{W})$ 
\be
\label{22}
a_0 * b_0 := \sigma \Big( \sigma^{-1}(a_0) \circ  \sigma^{-1}(b_0) \Big).
\ee
This $*$--product is natural and of the Vey type. 

Moreover,
applying the definition from (\ref{22}) to the $*$--product of functions depending only on spatial coordinates we see that their $*$-product is the usual pointwise multiplication of functions.
Provided that we
multiply two elements of the form
\[
\Big[(p_1)^{i_1}\ldots (p_n)^{i_n}f(q^1, \ldots, q^n)\Big]*\Big[(p_1)^{j_1}\ldots (p_n)^{j_n}g(q^1, \ldots, q^n)\Big],
\] 
we see
from   Corollary \ref{co2}  that the maximal power of $\hbar$ appearing in this product does not exceed $\hbar^{i_1+\cdots+i_n+j_1+ \cdots +j_n}.$

Hence in general $p_{\alpha} *p_{\beta} = p_{\alpha} \cdot p_{\beta} +\hbar^2 f,$
where the function $f$
 depends only on the coefficients  of the linear connection from the base space (see relation (\ref{nowa2})). Thus if the configuration space ${\cal M}$ is flat then we can always choose the spatial coordinates so that for all momenta canonically conjugated
with them  $p_{\alpha}*p_{\beta}=p_{\alpha} \cdot p_{\beta}.$

 Let us  write $\sigma^{-1}(a_0)[z]$ as
\[
\sigma^{-1}(a_0)[z_a]=a[z_a]= \sum_{d_a=0}^{\left[ \frac{z_a-1}{2}\right]}\hbar^{2d_a} a_{2d_a}[z_a-4d_a].
\]
Then for every $*$--product calculated
according to the Fedosov method the term $B_i(a_0,b_0)$ depends only on elements $a[z_a], b[z_b]$, which satisfy the following relations
\bea
\label{wielkanoc}
1 \leq z_a,z_b \leq 2i-1,\;&\; 
z_a+z_b=2i,\;&\; 
  d_a +d_b \leq \frac{i-1}{2}, \nonumber \\
z_b-z_a = 4(d_b -d_a),\;&\;
z_a-4d_a \leq i ,\;&\; z_b-4d_b \leq i .
\eea 

Assume that  the $*$-product is generated on a cotangent bundle  according to Fedosov's algorithm with some compatible symplectic connection. Then in a proper Darboux chart all of the expressions $B_i(a_0, b_0)$ for a fixed $i \geq 1$  are sums of elements
\be
\label{20032010-1}
f(q^1,\ldots, q^n) \frac{\partial^{\;k_1 + \ldots +k_{2n} }a_0}
{\partial(q^1)^{k_1} \ldots \partial(p_n)^{k_{2n}}}
\frac{\partial^{\;j_1 + \ldots +j_{2n}}b_0
}{\partial(q^1)^{j_1} \ldots \partial(p_n)^{j_{2n}}}
\;\;\; {\rm and } 
\ee
\be
\label{20032010}
p_{1}^{s_1}\ldots p_{n}^{s_n} g(q^1,\ldots, q^n) \frac{\partial^{\;k_1 + \ldots+ k_{2n}}a_0
}{\partial(q^1)^{k_1} \ldots \partial(p_n)^{k_{2n}}}
\frac{\partial^{\;j_1 + \ldots +j_{2n}}b_0
}{\partial(q^1)^{j_1} \ldots \partial(p_n)^{j_{2n}}},
\ee
\vspace{0.2cm}
\[
\exists_{1 \leq v \leq n}\; s_v>0,\;\;\; 
1 \leq k_1 + \cdots +k_{2n}\leq i \;\;\;, \;\;\; 1 \leq j_1 + \cdots +j_{2n}\leq i.
\]
From (\ref{kiedy}) we see that
$
\sum_{u=n+1}^{2n} k_u +\sum_{u=n+1}^{2n} j_u - i = \sum_{v=1}^{n}s_v.
$
Moreover, $\sum_{v=1}^{n}s_v \leq \left[\frac{i}{2}\right].$

Let us assume that $a_0(q^1,\ldots, q^n)$ is some function only of spatial coordinates. Then  the component
$B_i(a_0, b_0)$ of the
product of $a_0$  with an arbitrary function $b_0(q^1, \ldots,p_n)$ satisfies the following properties:
\begin{enumerate}
\item
$z_b - z_a\stackrel{\rm (\ref{wielkanoc})}{=}4d_b$ and therefore $z_b \geq i \geq z_a,$
\item
$z_b=i+ 2d_b$ so $z_a, z_b$ and $i$ are of the same parity,
\item
only components of $\sigma^{-1}(b_0)$ containing exclusively momenta-like $y^K$ are present in the product $a_0 * b_0.$ Thus the terms of the kind (\ref{20032010}) from the series $\sigma^{-1}(b_0)$ do not appear in the deformed multiplication.
\item 
Hence
$
\forall_{i\geq 1 } \;\; B_i(a_0, b_0)= 
\sum_{j_{n+1} + \ldots +j_{2n}=i}\;\; \sum_{1 \leq i_1 + \ldots +i_{n} \leq i}
g_{i_1 \ldots j_{2n}}(q^1,\ldots, q^n) \times
$
\[
\times
\frac{\partial^{\;i_1 + \ldots +i_{n}}a_0
}{\partial(q^1)^{i_1} \ldots \partial(q^n)^{i_{n}}} \;
\frac{\partial^{\;j_{n+1} + \ldots +j_{2n}}b_0
}{\partial(p_1)^{j_{n+1}} \ldots \partial(p_n)^{j_{2n}}}.
\]
\item
If $b_0=b_1(q^1,\ldots, q^n)\cdot b_2(p_1, \ldots, p_n)$ then
$
a_0 * b_0= b_1 \cdot (a_0 * b_2). 
$
\end{enumerate}

The canonical variables $q^1,\ldots, p_n$ fulfill commutation relations consistent with the Dirac quantization rules 
\be
\label{nowe10}
\{q^{\alpha},q^{\beta}\}_M=0 \;\;,\;\; 
 \{q^{\alpha},p_{\beta}\}_M=-i \hbar \delta^{\alpha}_{\beta} \;\;,\;\;
 \{p_{\alpha},p_{\beta}\}_M=0. 
\ee
By the symbol $\{ \cdot, \cdot \}_M$ we denote the Moyal bracket of functions
$
\{a_0,b_0\}_M := a_0*b_0-b_0*a_0.
$
Observe that the sign  in the second Equation (\ref{nowe10}) is the consequence of the Fedosov sign convention. Commutation  relations (\ref{nowe10}) are invariant under the point transformations (\ref{240709}). 

\subsection{Examples}

Assume the compatible symplectic connection is the induced symplectic connection defined by  relations (\ref{wzor1}). As we remember, in this situation the symplectic curvature tensor satisfies  property (\ref{e9}). Moreover, from the relation (\ref{e38}) we can see that this symmetry is inherited by the terms $R[0|i_1, \ldots, i_n|0|\alpha,\beta+n](y^1)^{i_1}\cdots (y^n)^{i_n} dq^{\alpha} \wedge dp_{\beta}$ of higher degrees. Thus we conclude that the Abelian connection series generated by the induced symplectic connection (\ref{wzor1}) is determined by the relatively simple formulas:
\[
\forall_{i_1, \ldots, i_n}\; \forall_{\alpha} \;\; r[0|i_1, \ldots, i_n|0|\alpha + n]=0,
\]
\[
\forall_{i_1, \ldots, i_n}\; \forall_{\alpha, \tau} \;\; r[0|i_1, \ldots, i_n|\tau|\alpha ]=- R[0|i_1, \ldots, i_n|0|\alpha, \tau +n ],
\]
\vspace{0.05cm}
\[
\forall_{i_1, \ldots, i_n}\; \forall_{\alpha, \beta} \;\;
R[0|i_1, \ldots, i_n|0|\alpha,\beta+n]= -
r[\beta|i_1, \ldots, i_n|0|\alpha]+
\] 
\be
\label{dzis}
 - \sum_{m=1}^n  \sum_{\stackrel{ g_1+\ldots +g_n=1}{\forall_c 0 \leq g_c \leq {\rm min}[i_c,1] }}(g_m+1)\cdot
\gamma[0|g_1, \ldots, g_m+1, \ldots,g_n|0|\beta+n]  \cdot R[0|i_1-g_1,\ldots, i_m-g_m, \ldots, i_n-g_n|0|\alpha, m+n].
\ee

Let us mention   the case  when the compatible  symplectic connection is of the form (\ref{pion1}). Then
the Abelian connection consists only of elements of the first and  second type. 
Thus every component $B_i(a_0, b_0)$  of the product $a_0 * b_0$ is only a sum of terms of the kind (\ref{20032010-1}). Functions
 $f(q^1, \ldots,q^n)$ from the formula (\ref{20032010-1}) are polynomials in the linear connection $\Gamma^{\alpha}_{\beta \delta}$  coefficients and their partial derivatives. These functions do not change along the fibres of the cotangent bundle ${\cal T^*M}$.
 
\section{Conclusions}

Starting from a few physical prerequisities 
we introduced some class of  symplectic connections
on a cotangent bundle ${\cal T^*M}$. 
These connections are constructed in a proper Darboux atlas and their construction is global. Since
they are modelled by a linear connection from the base space ${\cal M}$ we call them  symplectic connections compatible with a given linear connection.
There are many  symplectic connections compatible with the same linear symmetric connection. Their geometric description is contained in Theorem \ref{twfree}.

 It seems that among all symplectic connections with which the cotangent bundle ${\cal T^*M}$ can be equipped, the compatible symplectic connections are the most natural. The reasons for this statement are that they contain the linear connection from the base manifold ${\cal M }$ and are of simple structure. 

There is a deep geometrical relation between the  compatible symplectic connection and its original linear connection. For every smooth curve on  ${\cal T^*M}$ and every vector field $X$ transport parallel along this curve with respect to the compatible symplectic connection we obtain that the projected vector field $\pi_*(X)$ is parallel propagated along the projected curve on ${\cal M}$ with respect to the linear connection generating our symplectic connection. 

A symplectic connection on a symplectic space can be used, via the Fedosov construction, to introduce a natural $*$--product on this symplectic manifold. We propose a detailed analysis of this construction for a compatible symplectic connection. We prove that there are only three kinds of terms appearing in the Abelian connection $\gamma +r$ series. This simplicity of the Abelian connection series influences the $*$--product. We enumerate several properties of this family of  
 $*$--products. 
 
 We stress that all  star products generated by compatible symplectic connections are equivalent. However, in quantum mechanics the equivalent $*$--products yield different physical results. It is also probable that quantum mechanics involves more than one  $*$--product. Thus we do not restrict ourselves to   the induced symplectic connection, which is the most elegant from the mathematical point of view but propose the whole family of compatible symplectic connections.
 
 It seems that in the case of the $*$--product generated by a compatible symplectic connection it is possible to find some explicit noniterative formula defining this $*$--product. However, such a  problem is beyond  our interest in the context of the current article.
 \newline
 {\bf Acknowledgements}
 
 I am grateful to Prof. Maciej Przanowski for his interest in this paper and a lot of comments. I would like to thank  Prof. Bogdan Przeradzki who helped me  clarify  the proof of Theorem \ref{twfree}.

\appendix*

\section{The formula determining $R[\upsilon|i_1, \ldots, i_n|0|\alpha, \beta]$ }
\label{appA}

After simple but tedious calculations based on  the  Bianchi identity  (\ref{e20.1}) and the results
(\ref{31}),  (\ref{1203102}), (\ref{9032010})  we obtain 
\[
R[\upsilon|i_1, \ldots, i_n|0|\alpha, \beta] = \frac{\partial R[0|i_1, \ldots, i_n|0|\alpha, \upsilon+n]}{\partial q^{\beta}}
- \frac{\partial R[0|i_1, \ldots, i_n|0|\beta, \upsilon+n]}{\partial q^{\alpha}} +
\sum_{l=1}^n (s_{l}+1) \Bigg(
\]
\[
  \sum_{\stackrel{s_1+ \cdots +s_n=z-1}{\forall_c \,{\rm max}[i_c-1,0]\leq s_c \leq i_c}} \Big[
(i_{\beta } - s_{\beta}+1 )R[0|s_1, \ldots,s_{l}+1, \ldots, s_n|0|\alpha, \upsilon+n]
\gamma[0|i_1-s_1, \ldots, i_{\beta}-s_{\beta}+1, \ldots, i_n-s_n|0| l+n]+
\]
\[
-  
(i_{\alpha}-s_{\alpha}+1)R[0|s_1, \ldots,s_{l}+1, \ldots, s_n|0|\beta, \upsilon+n]
\gamma[0|i_1-s_1, \ldots, i_{\alpha}-s_{\alpha}+1, \ldots, i_n-s_n|0| l+n] \Big]+
\]
\[
 + \sum_{\stackrel{1 \leq s_1+ \cdots +s_n \leq z-2}{\forall_c\, 0\leq s_c \leq {\rm min}[z-2,i_c]}} \Big[
 (i_{\beta } - s_{\beta}+1 )
R[0|s_1, \ldots,s_{l}+1, \ldots, s_n|0|\alpha, \upsilon+n]
r[0|i_1-s_1, \ldots, i_{\beta}-s_{\beta}+1, \ldots, i_n-s_n|0| l+n]+
\]
\[
 - 
(i_{\alpha}-s_{\alpha}+1)
R[0|s_1, \ldots,s_{l}+1, \ldots, s_n|0|\beta, \upsilon+n]
r[0|i_1-s_1, \ldots, i_{\alpha}-s_{\alpha}+1, \ldots, i_n-s_n|0| l+n] \Big]+
\]
\[
 + \sum_{\stackrel{1 \leq s_1+ \cdots +s_n \leq z-2}{\forall_c \,0\leq s_c \leq {\rm min}[z-2,i_c]}}
\Big[
R[0|s_1, \ldots,s_{l}+1, \ldots, s_n|0|\beta, \upsilon+n]
R[0|i_1-s_1, \ldots, i_{l}-s_{l}, \ldots, i_n-s_n|0| \alpha, l+n]+
\]
\[
 - 
R[0|s_1, \ldots,s_{l}+1, \ldots, s_n|0|\alpha, \upsilon+n]
R[0|i_1-s_1, \ldots, i_{l}-s_{l}, \ldots, i_n-s_n|0| \beta, l+n]\Big]\Bigg)+
\]
\[
+\sum_{l=1}^n (i_l-s_{l}+1) \Bigg(
\]
\[
 \sum_{\stackrel{s_1+ \cdots +s_n=z-1}{\forall_c \,{\rm max}[i_c-2,0]\leq s_c \leq i_c}}
\Big[(s_{\beta}+1)
R[0|s_1, \ldots,s_{\beta}+1, \ldots, s_n|0|\alpha, l+n]
\gamma[0|i_1-s_1, \ldots, i_{l}-s_{l}+1, \ldots, i_n-s_n|0|\upsilon +n]+
\]
\[
 - (s_{\alpha}+1)
R[0|s_1, \ldots,s_{\alpha}+1, \ldots, s_n|0| \beta, l+n]
\gamma[0|i_1-s_1, \ldots, i_{l}-s_{l}+1, \ldots, i_n-s_n|0|\upsilon +n]\Big]+
\]
\[
+ 
 \sum_{\stackrel{1 \leq s_1+ \cdots +s_n \leq z-2}{\forall_c \,0\leq s_c \leq {\rm min}[z-2,i_c]}}
\Big[(s_{\beta}+1)
R[0|s_1, \ldots,s_{\beta}+1, \ldots, s_n|0|\alpha, l+n]
r[0|i_1-s_1, \ldots, i_{l}-s_{l}+1, \ldots, i_n-s_n|0|\upsilon +n] +
\]
\[
- 
 (s_{\alpha}+1)
R[0|s_1, \ldots,s_{\alpha}+1, \ldots, s_n|0| \beta, l+n]
r[0|i_1-s_1, \ldots, i_{l}-s_{l}+1, \ldots, i_n-s_n|0|\upsilon +n]\Big] \Bigg).
\]
It is assumed that $\sum_{l=1}^n i_l=z>2.$



\begin{thebibliography}{555}

\bibitem{MO49}
 J. E.  Moyal,  {\it Proc. Camb. Phil. Soc.} {\bf 45},  99 (1949).
 
 \bibitem{WY31} 
H. Weyl, {\it The Theory of Groups and Quantum Mechanics}, Methuen, London 1931 (reprinted  Dover, New York  1950).
\bibitem{WI32}
 E. P. Wigner,  {\it Phys. Rev.} {\bf 40}, 749 (1932).
 
 \bibitem{GW46}
H. J. Groenewold, {\it  Physica} {\bf 12}, 405 (1946).

\bibitem{baf}
F. Bayen, M. Flato, C. Fronsdal, A. Lichnerowicz, and D. Sternheimer, {\it Lett. Math. Phys.} {\bf 1}, 521 (1977). 

\bibitem{bay}
F. Bayen, M.Flato, C. Fronsdal, A. Lichnerowicz and D.  Sternheimer, {\it Ann. Phys. NY} {\bf 111}, 61 (1978); {\it Ann. Phys. NY} {\bf 111}, 111 (1978).

\bibitem{6}
B. Fedosov, {\it J. Diff. Geom.} {\bf 40}, 213 (1994).

\bibitem{7}
B. Fedosov, {\it Deformation Quantization and Index Theory}, Akademie Verlag, Berlin 1996.

\bibitem{biel}
P. Bieliavsky, M. Cahen, S. Gutt, J. Rawnsley and L. Schwachh\"{o}fer, arXiv:math/0511194v2.

\bibitem{ja5}
J. Tosiek, {\it Acta Phys. Pol. } {\bf B 38}, 3069 (2007).

\bibitem{ja6}
J. Tosiek, {\it J. Phys. Conf. Series} {\bf 128}, 012024 (2008). 

\bibitem{bour}
F. Bourgeois and M. Cahen, {\it J. Geom. Phys.} {\bf 30}, 233 (1999).

\bibitem{gut5}
M. Cahen and S. Gutt, {\it Lett. Math. Phys.} {\bf 6}, 395 (1982). 

\bibitem{bor}
M. Bordemann, N. Neumaier and S. Waldmann, {\it J. Geom. Phys.} {\bf 29}, 199 (1999).

\bibitem{miel}
B. Mielnik, private communication.

\bibitem{pleb1}
J. F. Pleba\'{n}ski, M. Przanowski and F. Turrubiates, {\it Acta Phys. Pol. } {\bf B 32}, 3 (2001).

\bibitem{gel}
I. Gelfand, V. Retakh and M. Shubin, {\it Adv. Math.} {\bf 136}, 104 (1998).

\bibitem{koba}
S. Kobayashi and K. Nomizu, {\it Foundations of Differential Geometry}, Interscience Publishers, New York 1963.

\bibitem{yano}
K. Yano and S. Ishihara, {\it Tangent and Cotangent Bundles}, Marcel Dekker Inc., New York 1973.

\bibitem{sgu}
S. Gutt and J. Rawnsley, {\it Lett. Math. Phys.} {\bf 66}, 123 (2003).

\bibitem{olga}
O. Kravchenko, arXiv:math. SG/0008157.

\bibitem{vais2}
I. Vaisman, {\it J. Math. Phys.} {\bf 43}, 283 (2002).

\end{thebibliography}
\end{document}